%% file: main.tex
\documentclass[pdflatex]{sn-jnl}

\usepackage{numprint}
\usepackage[utf8]{inputenc} 
\usepackage{csquotes}
\usepackage[T1]{fontenc}    
\usepackage{tikz}   
\usepackage{dsfont} 
\usetikzlibrary{quantikz2}
\usepackage{xcolor}
\hypersetup{colorlinks=true,urlcolor=teal,
            linkcolor=teal,citecolor=teal}
\usepackage{glossaries-extra}
\usepackage{listings} 
\usepackage[inline]{enumitem}
\usepackage{url}            
\usepackage{booktabs}       
\usepackage{amsfonts}       
\usepackage{amsmath,amssymb}
\usepackage{nicefrac}       
\usepackage{microtype}      
\usepackage{cleveref}       
\usepackage{graphicx}
\usepackage{subcaption} 
\usepackage{xspace}
\usepackage{tabularx}
\usepackage{tcolorbox}
\usepackage{siunitx}

\usepackage{stmaryrd} 
\SetSymbolFont{stmry}{bold}{U}{stmry}{m}{n} 

\usepackage[
  backend=biber,
  maxcitenames=3,
  minbibnames=3,
  mincitenames=1,
  maxbibnames=3,
  uniquelist=false,
  url=false,
  doi=true,
  eprint=false,
  isbn=false,
]{biblatex}
\def\subfigurewidth{0.48\textwidth}

\input{commands.tex}

\begin{document}
\input{acronyms.tex}

\title[Out of Tune: Demystifying Noise-Effects on Quantum Fourier Models]{Out of Tune: Demystifying Noise-Effects on Quantum Fourier Models}

\author[1]{\fnm{Maja} \sur{Franz}
  \email{maja.franz@othr.de}}
\equalcont{These authors contributed equally to this work.}
\author[2]{\fnm{Melvin} \sur{Strobl}  \email{melvin.strobl@kit.edu}}
\equalcont{These authors contributed equally to this work.}

\author[2]{\fnm{Leonid} \sur{Chaichenets}}
\author[2]{\fnm{Eileen} \sur{Kuehn}}
\author[2]{\fnm{Achim} \sur{Streit}}
\author[1, 3]{\fnm{Wolfgang} \sur{Mauerer}}

\affil[1]{\orgname{Technical University of Applied Sciences Regensburg}, \orgaddress{\country{Germany}}}
\affil[2]{\orgname{Karlsruhe Institute of Technology}, \orgaddress{\country{Germany}}}
\affil[3]{\orgname{Siemens AG, Technology}, \orgaddress{\city{Munich}, \country{Germany}}}

\abstract{%
  Variational quantum algorithms have received substantial theoretical and empirical attention.
  As the underlying \glspl{vqc} can be represented by Fourier series that contain an exponentially large spectrum in the number of input features, hope for quantum advantage remains.
  Nevertheless, it remains an open problem if and how \glspl{qfm} can concretely outperform classical alternatives, as the eventual sources of non-classical computational power (for instance, the role of entanglement) are far from being fully understood.
  Likewise, hardware noise continues to pose a challenge that will persist also along the path towards fault tolerant quantum computers.

  In this work, we study \glspl{vqc} with Fourier lenses, which provides possibilities to improve their understanding, while also illuminating and quantifying constraints and challenges.
  We seek to elucidate critical characteristics of \glspl{qfm} under the influence of noise.
  Specifically, we undertake a systematic investigation into the impact of noise on the Fourier spectrum, expressibility, and entangling capability of \glspl{qfm} through extensive numerical simulations and link these properties to training performance.
  The insights may inform more efficient utilisation of quantum hardware and support the design of tailored error mitigation and correction strategies.

  Decoherence imparts an expected and broad detrimental influence across all \ase.
  Nonetheless, we observe that the severity of these deleterious effects varies among different model architectures, suggesting that certain configurations may exhibit enhanced robustness to noise and show computational utility.
}


\maketitle
\glsresetall

\section{Introduction}

Given the remarkable advancements in machine learning, significant optimism has been directed towards \gls{qml}.
However, such enthusiasm has frequently been tempered by the prevailing limitations of contemporary quantum hardware~\cite{Greiwe:2023}, together with algorithmic shortcomings~\cite{schuld22_qadvantage}.
The exact capabilities of \gls{qml} are not yet fully understood, but considerable opportunities to achieve advantage over classical approaches remain.
As the concept of \gls{qml} (see the introductory review by \citeauthor{schuld_supervised_2018}~\cite{schuld_supervised_2018}) has received considerable attention in the literature during the last years, crucial limitations and trade-offs have been identified~\cite{zimboras_myths_2025}, especially regarding trainability challenges~\cite{ragone_lie_2024} and model complexity constraints~\cite{heimann_learning_2024}.
Despite the progress from \gls{nisq} devices~\cite{georgopoulos_modeling_2021} towards \gls{ftqc}, noise and imperfections will continue to influence \gls{qml} performance and properties in the foreseeable future, even when considering the best possible use of already available hardware resources~\cite{Wintersperger:2022}.

In this work, we aim to shed light on the impact of noise on key properties of \glspl{vqc}, the core computational resource of \gls{qml}.
Specifically, we focus on \glspl{qfm}, a representation of \glspl{vqc} as truncated Fourier series.
We emphasise that our results are hardware agnostic, that is, they are not only applicable to error-corrected architectures, but also to intermediate \glspl{qpu} and algorithms (\eg as in Refs.\cite{Thelen:2023, Franz:2023, Franz:2024, Periyasamy:2024, Bayerstadler:2021, Carbonelli:2024, Gogeissl:2024}), until full \glspl{ftqc} are available.

Common metrics used to assess \glspl{vqc} include expressibility and entangling capability~\cite{sim_expressibility_2019}, which generally determine how effectively a \gls{vqc} can explore the Hilbert space.
This exploration is crucial because it indicates the capacity of a \gls{vqc} to learn various functions, with the solutions to these functions residing in this Hilbert space.
However, a highly expressive \gls{vqc} can face challenges such as the \gls{bp} problem~\cite{ragone_lie_2024}, where the optimisation landscape becomes difficult to navigate.
Therefore, a successful \gls{vqc} must strike a balance: It needs to be sufficiently expressive so that the portion of the Hilbert space it accesses contains potential solutions, while maintaining a manageable optimisation landscape.
Another way to gauge this property is by examining the number and values of the \gls{qfm}'s Fourier coefficients~\cite{mhiri_constrained_2024}, which dictate the types of functions the \gls{qfm} can learn.
Conversely, if an \as lacks sufficient expressibility, the \emph{trainable Hilbert space} it accesses is highly unlikely to contain viable solutions.

\emph{Entangling capability} is another metric to determine the amount of entanglement a \gls{vqc} can create using optimal parameters.
This metric can be viewed as the \emph{quantumness} a particular circuit can achieve and is therefore crucial to set it apart from classical surrogates or \glspl{vqc} without entanglement.

In this work, we determine the Fourier spectral properties, expressibility, and entangling capability of commonly used \ase under the influence of noise using extensive numerical simulations.
Furthermore we link these characteristics to the performance of the \glspl{qfm} in machine learning tasks by conducting trainings on synthetically generated regression problems.

\begin{figure}[tbp]
  \centering
  \begin{subfigure}[b]{\subfigurewidth}
    \includegraphics[clip, trim=0.8cm 0.1cm 0.4cm 0.2cm, width=\columnwidth]{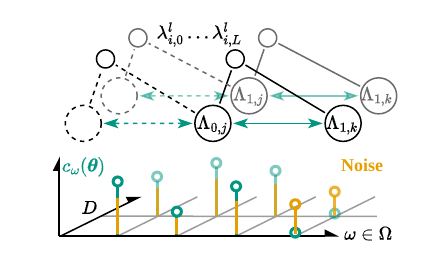} 
    \caption{}
    \label{fig:overview-a}
  \end{subfigure}\hfill%
  \begin{subfigure}[b]{\subfigurewidth}
    \includegraphics[clip, trim=0.3cm 0cm 0cm 0.3cm, width=\columnwidth]{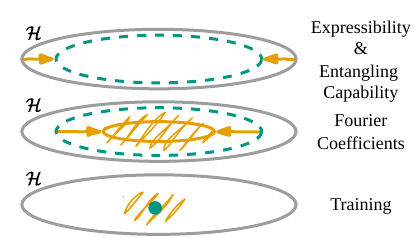}
    \caption{}
    \label{fig:overview-b}
  \end{subfigure}
  \caption{(a) Quantum spectrum tree~\cite{mhiri_constrained_2024} where $\Lambda$ is constituted by the sum of the individual eigenvalues $\lambda$ of each generator in the $D$-dimensional input encoding.
    The spectrum $\Omega$ is determined by the differences between each $\Lambda$ where each coefficient \textcolor{lfd4}{$c_\omega(\btheta)$} depends on a parameter vector $\btheta$.
    \textcolor{lfd2}{Noisy coefficients} generally have a reduced magnitude, which can lead to non-trainable frequency components.
    (b) Implications of the left considerations on Hilbert space level (solid grey), denoted by $\mathcal H$: Expressibility and entangling capability limits depend on the level of \textcolor{lfd2}{noise (arrows)} and the \textcolor{lfd4}{accessible space (dashed)} of the model.
    Additionally, noise further reduces this space when viewed through the Fourier lens, as it reduces the ability to change frequency components while adding untrainable \textcolor{lfd2}{frequency artifacts} (hatched outside of solid orange).
    Similarly, in a training scenario, \textcolor{lfd2}{noise} causes a shift away from the \textcolor{lfd4}{target space} as frequency components get affected differently by noise.}
  \label{fig:overview}
\end{figure}

Our approach aims to provide a nuanced perspective on the factors that contribute to the aptitude of \gls{qml} for specific learning tasks.
The main contributions of this work are illustrated in~\autoref{fig:overview}.
Additionally, we provide a consistent collection of notation on the \gls{qfm} characteristics, together with a software reproduction package~\cite{franz_reproduction_2025} that encompasses the computation of the relevant metrics using our \emph{QML-Essentials} Python framework~\cite{strobl2025qmlessentialsframework}.

The rest of this paper is structured as follows: Before we review related work in~\autoref{sec:related_work}, we introduce the concept behind \glspl{qfm} in~\autoref{sec:qfm}.
We then present our methods in~\autoref{sec:method}, where we introduce the noise models used in this work, discuss how we measure coefficients, entangling capability and expressibility, and define a synthetic dataset on which we train the \glspl{qfm}.
\autoref{sec:results} presents our main contribution, which is the numerical results on the computation of the aforementioned characteristics in \glspl{vqc}.
Here we measure Fourier coefficients, expressibility, and entangling capability in a quantitative manner through random parameter sampling, and also in a synthetic training setting.
In \autoref{sec:discussion} we provide a discussion and finally conclude in~\autoref{sec:conclusion}.

\clearpage 

\section{Quantum Fourier Models}
\label{sec:qfm}

We consider an $n$-qubit quantum circuit whose expectation value, measured on each qubit ($n$-local) using the observable $\mathcal{M}$, is given by
\begin{equation}
  f(\bx, \btheta)=\bran{0} U^{\dagger}(\bx, \btheta) \mathcal{M} U(\bx, \btheta)\ketn{0},
  \label{eq:circuitExpr}
\end{equation}
which is parametrised by the $D$-dimensional input $\bx = (x_1, \dots, x_D) \in \mathbb{R}^D$ and the trainable parameter vector $\btheta = (\btheta_1, \dots, \btheta_{L+1}) \in [\mathbb{R}^K]^{L+1}$ with $L$ layers of $K$ parameters per layer\footnote{Without loss of generality, we choose the same number of parametrised gates for each layer.}.

We build the eventual unitary $U(\bx, \btheta)$ as products of multiple layers ($\ell \in [1,L]$) of consecutive encoding unitaries $S^{(\ell)}(\bx)$ and trainable unitaries $W^{(\ell)} \coloneqq W^{(\ell)}(\btheta_\ell)$ as follows:
\begin{equation}
  U(\bx)=W^{(L+1)} S^{(L)}(\bx) W^{(L)} \cdots W^{(2)} S^{(1)}(\bx) W^{(1)}
\end{equation}
where an additional last trainable layer $W^{(L+1)}$ is added.
To allow for an arbitrary circuit structure and placement of unitary and encoding gates, we construct $S^{(\ell)}(\bx)$ to comprise combinations of $D$ unitaries $S^{(\ell)}_i(\bx)$, each of which encodes one single input feature $x_i$:
\begin{equation}
  S_i^{(\ell)}(\bx) = \exp{(-\I \be_i^T \bx G_i^{(\ell)})} = \exp{(-\I x_i G_i^{(\ell)})}
\end{equation}
with Hermitian generator $G_i^{(\ell)}$ and standard basis vector $\be_i$.

As shown in the seminal contributions in Refs.~\cite{schuld_effect_2021,perez-salinas_data_2020}, such an architecture allows for rewriting~\autoref{eq:circuitExpr} as a partial Fourier series
\begin{equation}
  \begin{aligned}
    f(\bx, \btheta) & = \sum_{\bomega \in \bOmega} c_{\bomega}(\btheta) e^{\I \bomega^T \bx} \\
                    & = \sum_{\bomega \in \bOmega}
    \vert c_{\bomega}(\btheta) \vert
    \left(\cos(\varphi(\btheta)) \cos(\bomega^T \bx) - \sin(\varphi(\btheta) \sin(\bomega^T \bx)\right),
  \end{aligned}
  \label{eq:qfm}
\end{equation}
where $\bOmega$ contains the unique frequencies resulting from the eigenvalues of each generating encoding Hamiltonian $G_i^{(\ell)}$.
The corresponding complex Fourier coefficients are given by $\{c_{\bomega}(\btheta)\} = \{\vert c_{\bomega}(\btheta) \vert e^{\I \varphi(\btheta)}\}$.

Following the definition in Ref.~\cite{mhiri_constrained_2024}, the resulting set of frequencies is
\begin{equation}
  \bOmega=\left\{ \times_{i=1}^{D} \left\{\Lambda_{i, \boldsymbol{j}}-\Lambda_{i, \boldsymbol{k}}\right\} \mid \boldsymbol{j}, \boldsymbol{k} \in \llbracket 1, d_\ell \rrbracket^L_{\ell=1}\right\},
  \label{eq:set_of_frequencies}
\end{equation}
where the bracket $\llbracket \cdot \rrbracket$ defines the product of the intervals $[ \cdot ]$, s.t. $\boldsymbol{j}$ (and $\boldsymbol{k}$) is a multi-index $(j_1, j_2, \dots, j_L)$ with each $j_\ell \in [1, d_\ell]$.
The corresponding $\Lambda_{\boldsymbol{j}}$ presents the sum of all $d_\ell$ eigenvalues of the $\ell$-th generator grouped across all layers, given by
\begin{equation}
  \Lambda_{i, \boldsymbol{j}}=\lambda_{i,j_1}+\cdots+\lambda_{i,j_L}.
\end{equation}
Here,~\autoref{eq:set_of_frequencies} restates a finding from Ref.~\cite{casas_multidimensional_2023} where it was shown that the set of frequencies is constituted by the Cartesian product over all the individual sets that stem from each input encoding.
While there are no constraints on the position of each input encoding unitary inside the circuit, it is important to note that the useable set of frequencies only extends if all data encoding unitaries are a mutually orthogonal set of functions.

Another descriptive metric of the Fourier spectrum that also provides an upper bound for the expressibility was introduced in Ref.~\cite{mhiri_constrained_2024} in form of the frequency redundancy which is defined by the size of the redundancy generator
\begin{equation}
  R({\bomega}) =
  \left\{(\boldsymbol{j}, \boldsymbol{k}) \mid \times_{i=1}^{D} \left\{\Lambda_{i, \boldsymbol{j}}-\Lambda_{i, \boldsymbol{k}}\right\} = \bomega \land \boldsymbol{j}, \boldsymbol{k} \in \llbracket 1, d_\ell \rrbracket^L_{\ell=1} \right\}.
\end{equation}
Considering single qubit Pauli-encodings, by increasing the number of layers, the number of distinct frequencies $\lvert \bOmega \rvert$ grows linearly, whereas the number of redundancies $\sum_{\bomega \in \bOmega} \lvert R(\bomega) \rvert$ grows exponentially.
Notably, it was also shown in Ref.~\cite{schuld_effect_2021}, that the exact same result can be achieved by increasing the number of qubits instead, such that $\vert \bOmega \vert \sim (nL)^D$.
Generally, the increasing number of frequencies stem from the gaps between the eigenvalues of the feature map generating Hamiltonians as visualised in~\autoref{fig:overview-a}.

The coefficients $c_{\bomega}(\btheta)$, parametrised by the trainable parameters in the \gls{vqc}, are solely determined by the circuit architecture (also including the input encoding structure~\cite{caro_encoding-dependent_2021}).
The way parameters act on the coefficients is non trivial with an analytical description of this relation as published in Refs.~\cite{nemkov_fourier_2023, wiedmann_fourier_2024}.

\section{Related Work}
\label{sec:related_work}

Considering noise in the context of \glspl{qfm} involves a multitude of relevant aspects, ranging from \emph{Fourier analysis} through \emph{trainability}, \emph{dequantisation}, \emph{expressibility}, and \emph{entanglement}.
In this section, we review the main body of relevant related work.

The seminal work of \citeauthor{perez-salinas_data_2020}~\cite{perez-salinas_data_2020} introduced the \emph{data reuploading} technique that enables a non-linear transformation between input and output of a \gls{vqc} with unitary transformations.
A follow-up study by \citeauthor{schuld_effect_2021}~\cite{schuld_effect_2021} sparked a whole subfield of research within \gls{qml} by establishing how the spectrum of a \gls{qfm} changes with the input encoding.
\citeauthor{casas_multidimensional_2023}~\cite{casas_multidimensional_2023} extended the idea to multidimensional input features and found that for some \ase, the size of the spectrum grows faster than the available degrees in a Hilbert space.
A fixed encoding strategy, as used in the aforementioned papers, results in an evenly spaced spectrum. \citeauthor{jaderberg_let_2024}~\cite{jaderberg_let_2024} found that by adding a trainable parameter to the input, distances between the frequencies can be modified.
The expressibility of \gls{qfm} and the redundancies of eigenvalues that effectively determine the resulting spectrum was linked by \citeauthor{mhiri_constrained_2024}~\cite{mhiri_constrained_2024}, which also demonstrated that the variance of a frequencies coefficients linearly depends on the number of its redundancies.
\citeauthor{nemkov_fourier_2023}~\cite{nemkov_fourier_2023} introduced an analytical description between Fourier coefficients and trainable parameters.
\citeauthor{wiedmann_fourier_2024}~\cite{wiedmann_fourier_2024} use this analytical description to argue that certain coefficients (frequencies) can vanish depending on the trainable parameters, which leads to a reduction of the accessible spectrum.

While there has been a significant effort to build mathematical formulations and find analytical relations around \glspl{qfm}, the trainability of such models has to be regarded as well.
One of the fundamental problems is the \glsxtrfull{bp} phenomenon, introduced in Ref.~\cite{cerezo_cost_2021}.
Generally, it describes the exponential decay of gradients in a \gls{vqc} caused by
\begin{enumerate*}[label=(\roman*)]
  \item the circuit expressibility;
  \item degree of entanglement;
  \item locality of the observable~\cite{ragone_lie_2024, larocca_barren_2025} and
  \item noise~\cite{wang_noise-induced_2021, ragone_lie_2024}.
\end{enumerate*}

Simply avoiding \glspl{bp} might look like an obvious solution, but it is known that circuits without a \gls{bp} can be classically simulated, eliminating any potential quantum advantage~\cite{cerezo_does_2023}  for a majority of use cases.
Also, Ref.~\cite{caro_encoding-dependent_2021} showed that the encoding strategy plays a significant role in the trainability of \gls{qml} models in terms of avoiding \glspl{bp}.

To assess whether \gls{qml} offers advantages over classical \gls{ml}, one approach is to employ dequantisation techniques that seek to understand if classical alternatives constructed from a quantum circuit can match or surpass \gls{qml} performance.
\citeauthor{schreiber_classical_2022}~\cite{schreiber_classical_2022} demonstrated that such surrogate models can outperform \glspl{qfm} for small problem instances, but this becomes intractable as the number of Fourier coefficients grows exponentially with input features.
To address this issue, \citeauthor{fontana_classical_2023}~\cite{fontana_classical_2023} and \citeauthor{landman_classically_2022}~\cite{landman_classically_2022} proposed approximating \gls{vqc} outcomes by  trimming frequencies or using \glspl{rff}.
\citeauthor{sweke_potential_2025}~\cite{sweke_potential_2025} established that efficient dequantisation for regression is possible with \glspl{rff} if the spectrum of the \gls{qfm} is polynomially concentrated, requiring a polynomial number of frequencies.

The aforementioned references consider \gls{vqc}, optimisation algorithm and training data as a whole, which is reasonable from an end-to-end point of view of \gls{qml}.
However, this holistic way of considering \gls{qml} makes it hard to assess the influence of the actual \gls{vqc} \emph{structures} that are the core computational resources.
%
\label{sec:entanglemnt_rel_work}
Characterising the properties of \glspl{vqc} to identify suitable \ase for variational quantum algorithms is a common approach, while the implications are often left as open questions in the field.
In this context, \citeauthor{sim_expressibility_2019}~\cite{sim_expressibility_2019} conducted an extensive study to measure the expressibility (\cf~\autoref{sec:expressibility}) and entangling capability (\cf~\autoref{sec:entangling_capability}) of different \ase.
Expressibility, which indicates how effectively an \as can explore the Hilbert space, has known relations to redundancies in the Fourier spectrum~\cite{mhiri_constrained_2024}.

The role of entanglement in \gls{qml} and other variational algorithms remains far from fully understood.
When dealing with input data that is inherently quantum in nature, entanglement is often considered as the key resource for successful learning~\cite{sharma_reformulation_2022}.
\citeauthor{wang_noise-induced_2021}~\cite{wang_transition_2024} demonstrated that increasing entangling capability up to a certain threshold can lead to improved model performance.
However, when the input data is classical, empirical studies, such as those in Refs.~\cite{bowles_better_2024, rohe_questionable_2024}, have shown that a low-entanglement circuit can perform similarly well, or even better than highly entangled ones on specific learning tasks.
Furthermore, \citeauthor{joch_entanglement-informed_2025}~\cite{joch_entanglement-informed_2025} highlighted the potential downside of excessive entanglement in learning scenarios, suggesting that high entangling capability can lead to the \gls{bp} phenomenon.
This indicates a critical threshold beyond which entanglement may no longer be beneficial for learning tasks.
To our knowledge, there is currently no established connection in previous work between the entangling capability of a circuit and its Fourier spectrum, to which we contribute in this work.


Given that fault-tolerance is not yet fully in reach (and does not automatically imply zero error), the influence of noise on \glspl{qfm} remains an important open question.
\citeauthor{Dalzell_2024}~\cite{Dalzell_2024} showed that on random quantum circuits, weak local noise translates into global white noise after measurement.
More precisely, incoherent local errors get scrambled by random quantum circuits causing the output to approach a uniform distribution.
\citeauthor{fontana_spectral_2022}~\cite{fontana_spectral_2022} measured the effect of noise on the Fourier spectrum to make suggestions for noise mitigation and diagnostics.
By cutting or filtering certain noise-induced coefficients or frequencies, noiseless landscapes can be approximately reconstructed.
The authors focus on the \gls{qaoa} and \gls{vqe}, for which they derive a uniform contraction of the Fourier coefficients when applying certain types of decoherent noise.
Apart from Ref.~\cite{fontana_spectral_2022}, we are not aware of other publications that investigate the direct influence of noise on the Fourier spectrum and other properties of \glspl{vqc}, especially in the \gls{qml} domain.
With this article, we aim to contribute to filling this research gap.

\section{Method}\label{sec:method}

In the following, we present our methods, which encompass the noise models employed, along with the notations and metrics utilised to assess the characteristics of \glspl{qfm}. These characteristics include Fourier coefficients, expressibility, and entangling capability. Additionally, we detail the synthetic dataset constructed for training these \glspl{qfm}.

\subsection{Noise}
\label{sec:noise}
We follow a standard approach of modelling decoherent
noise~\cite{georgopoulos_modeling_2021,Greiwe:2023,maschek:25:noise} by three mechanisms,
namely
\begin{enumerate*}[label=(\roman*)]
  \item damping noise (\ie, environmental effects, decoherence) encompassing \gls{ad} and \gls{pd};
  \item decoherent gate errors, including \gls{bf}, \gls{pf} and \gls{dp} noise and
  \item \gls{spam} errors.
\end{enumerate*}
In addition to the aforementioned, we also consider \glspl{cge}.
Details on the modelling of the different noise types are available in~\autoref{apx:noise}.

\subsection{Coefficients}

Based on the definition of a \gls{qfm} from~\autoref{eq:qfm}, we investigate the coefficients of such a model after parametrisation using the \gls{fft}.
Given the correct number of frequencies, which is in the simplified scenario of Pauli-encoding and a single encoding layer equal to the number of qubits $n$, this transformation yields a numerical estimate of the coefficients.
Note that this set of coefficients is a numerical approximation based on the expectation value of the model given a range of input samples $x\in\mathcal{X}$ and a fixed parameter vector $\btheta$.
As stated in~\autoref{eq:qfm}, these coefficients depend on the parameters.
For each \as we assume that this set of coefficients is characterised by the mean and the standard deviation which is what we evaluate in the following numerical experiments.

We calculate the covariance matrices as
\begin{equation}
  \Cov_{c} = \left(\begin{smallmatrix}
      \Cov(\Real(c), \Real(c)) & \Cov(\Real(c), \Imag(c)) \\
      \Cov(\Imag(c), \Real(c)) & \Cov(\Imag(c), \Imag(c))
    \end{smallmatrix}\right).
  \label{eq:coeff_covar}
\end{equation}

To estimate the model spectrum, we sample the parameter space $\Theta$.
We estimate the mean value of a coefficient contributing to the frequency $\bomega$,
\begin{equation}
  \mu_c(\bomega) = \frac{1}{\vert \Theta\vert}\sum_{\btheta\in\Theta} \left\vert\mathtt{FFT}_{\mathcal{X}}(f(\cdot, \btheta))(\bomega)\right\vert,
  \label{eq:coeff_fft_mu}
\end{equation}
and the relative standard deviation
\begin{equation}
  \sigma_c(\bomega) = \frac{1}{\vert \Theta\vert}\sum_{\btheta\in\Theta} \frac{\sqrt{\left\vert\mathtt{FFT}_{\mathcal{X}}(f(\cdot, \btheta))(\bomega)-\mu_c(\bomega)\right\vert^2}}{\mu_c(\bomega)}.
  \label{eq:coeff_fft_sigma}
\end{equation}
$\mathtt{FFT}_\mathcal{X}$ represents the discrete Fourier transform over $x\in\mathcal{X}$.
Analytical coefficients can be obtained by expanding the expectation value using trigonometric polynomials~\cite{nemkov_fourier_2023, wiedmann_fourier_2024}.
Prior to our numerical experiments, we use the analytical method to cross-validate the results obtained by the \gls{fft}.
This is to ensure a sufficient number of frequencies that has to be estimated in advance to satisfy the Shannon-Nyquist sampling theorem, \ie $\vert \mathcal{X}\vert \ge 2\max(\Omega)$.

\subsection{Expressibility}
\label{sec:expressibility}
To determine expressibility, we utilise the \gls{kl} divergence between the distributions
\begin{enumerate*}[label=(\roman*)]
  \item obtained by sampling from the Haar integral $\int_{\text {Haar }}\text{d} \psi (|\psi\rangle\langle\psi|)^{\otimes t} $ of a state $t$-design and
  \item obtained from the model $\int_{\btheta}\text{d}\btheta \left(\left|\psi_{\btheta}\right\rangle\left\langle\psi_{\btheta}\right|\right)^{\otimes t}$~\cite{sim_expressibility_2019,kullback_information_1951}:
\end{enumerate*}
\begin{equation}
  D_{\mathrm{KL}}\left(\hat{P}_{\text{Model}}(F ; \btheta) \| P_{\text {Haar }}(F)\right).
  \label{eq:kl_divergence}
\end{equation}
Here, the fidelity $F=\left|\left\langle\psi_{\boldsymbol{\varphi}} \mid \psi_{\boldsymbol{\phi}}\right\rangle\right|^2$ is based on the state overlap whereas the distributions of state overlaps is then $p\left(F\right)$.

This metric yields zero if $\hat{P}_{\text{Model}}(F ; \btheta) = P_{\text {Haar }}(F)$, meaning the states sampled from the \gls{qfm} are Haar distributed.
For the least expressive case, that is, the empty circuit $U=\mathds{1}$, the KL divergence becomes $\ln(n_\text{bins})$ where $n_\text{bins}$ describes the number of bins that are used for discretising the probability distribution using a histogram.
For the remainder of this work, we refer to the expressibility as the inverse of \gls{kl} divergence.

\subsection{Entangling Capability}
\label{sec:entangling_capability}
There are different ways to calculate the entangling capability of a \gls{vqc} (as there are many different ways of quantifying entanglement~\cite{plenio_introduction_2006}).

\subsubsection{Meyer-Wallach Measure}
\label{sec:entangling_capability_mw}

The \gls{mw} entangling capability~\cite{meyer_global_2002, brennen_observable_2003} of a state $\ket{\psi}$ is defined as
\begin{equation}
  Q(\ket{\psi})=2\left(1-1 / n \sum_{k=0}^{n-1} \operatorname{Tr}\left[\rho_{k}^{2}\right]\right),
  \label{eq:entangling_capability_brennen}
\end{equation}
based on partial density matrices $\rho_k$, where $k$ denotes the subsystem (\ie, qubit index).

This metric has the property that if $\operatorname{Tr}\left[\rho_{j}^{2}\right]=1 \quad \forall j$, implying $Q=0$, $\ket{\psi}$ is a product state whereas $Q=1$ iff $\operatorname{Tr}\left[\rho_{k}^{2}\right]=1 / 2 \quad \forall k$ meaning the state is maximally mixed. Notable, this metric is restricted to pure states, but cannot be used for mixed states as they occur in decoherent noisy circuits (despite working with
mixed states, we need the measure nonetheless, as will become clear in the next section).

\subsubsection{Entanglement of Formation}
\label{sec:entangling_capability_ef}

The \gls{ef}~\cite{wootters_entanglement_1998} can be used as a metric on mixed states. Following Ref.~\cite{plenio_introduction_2006}, it is usually defined as
\begin{equation}
  Q(|\psi\rangle):=\inf \left\{\sum_{i} p_{i} E\left(\left|\psi_{i}\right\rangle\left\langle\psi_{i}\right|\right): \rho=\sum_{i} p_{i}\left|\psi_{i}\right\rangle\left\langle\psi_{i}\right|\right\},
  \label{eq:entangling_capability_ef}
\end{equation}
and represents the average entanglement over all pure state decompositions of a density matrix $\rho$.
The entanglement for pure states is then calculated using the \gls{mw} measure as introduced in~\autoref{eq:entangling_capability_brennen}.
Finding a pure state decomposition however is a non-trivial task.
In this work we utilise an eigenvalue and eigenstate decomposition of the density matrix $\rho$.
Based on the resulting decomposition, we proceed with calculating the entanglement of each eigenstate while weighting it by its eigenvalue as depicted in the left part of the right hand side of~\autoref{eq:entangling_capability_ef}.

\subsection{Synthetic Training Dataset}
\label{sec:dataset}

To analyse the behaviour of \glspl{qfm} during optimisation, we employ random Fourier series as objective functions for regression tasks:
\begin{equation}
  f'(\bx) = \sum_{\bomega \in \bOmega} c'_{\bomega} e^{\I \bomega^T \bx} \quad \text{with} \quad c'_{\bomega} = ar_{\bomega}{\left|\sum_{\bomega \in \bOmega} r_{\bomega} e^{(\I \bomega^T \bx)}\right|^{-1}}.
\end{equation}
Here, normalisation factor $a$ and $r_{\bomega} \sim \text{Uniform}([0,1))$ are uniformly randomly sampled complex (or real-valued for $\bomega = \boldsymbol{0}$) numbers that satisfy the symmetry constraints of a real-valued Fourier series.
The normalisation factor, set to $a=0.5$ in our experiments, restricts the values of $f'(\bomega)$ to the interval $[-a, a]$, such that they are attainable by the \gls{qfm}.
The number of frequencies in the objective Fourier series is set to match the number of expected frequencies in the \gls{qfm} $\lvert \bOmega \rvert$ and the input data consists of evenly spaced points $\bx = (x_1, \dots, x_s)$ in the interval $[-\pi, \pi]$, where $s = \lvert \bOmega \rvert$.
While we only focus on one-dimensional input data ($D=1$) in the training experiments of this work, the approach can be extended to also generating higher-dimensional training datasets.

To evaluate the performance of a \gls{qfm} $f(\bx, \btheta)$, we use \gls{mse}
\(
\text{MSE}(f', f) = \frac{1}{s} \sum_{i=1}^s \left(f'(x_i) - f(x_i, \btheta)\right)^2\) as metric.
Additionally, we introduce a difference metric $\diffw$ to assess how closely a \gls{qfm} approximates the exact Fourier coefficients of the target function for each $\bomega$:
\begin{equation}
  \diffw = \lvert \lvert c_{\bomega}(\btheta)\rvert - \lvert c'_{\bomega}\rvert \rvert.
\end{equation}

\section{Numerical Results}
\label{sec:results}

We consider a circuit with the structure introduced in~\autoref{eq:circuitExpr} with $L = 1$ layers and $n \in [3\dots 6]$ qubits.
In the coefficient experiments, where we sample parameters, $D \in \{1, 2\}$ input features are encoded, while $D=1$ in the training experiments.
By convention, such a model has two trainable layers $U(\bx)=W^{2} S^{1}(\bx) W^{1}$.
However, for calculating expressibility and entanglement we discard encoding gates and the second trainable layer to make the results more consistent with Ref.\cite{sim_expressibility_2019}.
For the trainable part, we investigate the \ase depicted in~\autoref{fig:ansaetze} of \autoref{apx:ansaetze}.
We use the \gls{sea} as introduced in Ref.~\cite{schuld_circuit-centric_2020} and a circular structure for the \gls{hea}.
Both of these \ase are commonly-used within the \gls{qml}-community: The \gls{sea} presents an \as that utilises comparatively many entangling gates and trainable parameters; the \glspl{hea} uses fewer gates and its depth only grows as a constant with the number of qubits, which is a desirable property for \gls{nisq} circuits.
\gls{c15} and \gls{c19} stem from Ref.~\cite{sim_expressibility_2019} and are chosen based on their different structural properties (\cf \autoref{fig:coefficients_real_imag_encoding}) and distinguishable expressibility and entangling capability.
In particular, \gls{c15} presents an \as that only uses one type of rotation gate, which is potentially easier to implement on quantum hardware. \gls{c19} is chosen as a representative of \ase that use controlled rotation gates as an entangling gate, instead of CNOT gates, which may provide an additional degree of variability in the entangling capability (\cf \autoref{sec:entanglement}).

In the following experiments, we apply the types of noise described in~\autoref{sec:noise} to each \as with seven linearly increasing noise levels from 0\% to 3\% in steps of 0.5\%.
The trainable parameters are randomly uniformly sampled from $[0 \dots 2\pi]$.
For each parameter sample, we perform $n$ local measurements on all qubits and then collect and average the expectation value over all samples.
To ensure stability against parameter initialisation, we use different seeds upon model initialisation.
Throughout all our experiments, we use \num{250} samples per parameter value for each seed, which is scaled exponentially with the number of qubits if not stated otherwise (\eg, for a three qubit circuit with \num{10} parameters, we sample $10 \times 250 \times 2^{3}$ times for each seed).
While we use ten different initialisation seeds for the training, all other experiments in \autoref{sec:coefficients_exp}, \autoref{sec:expressibility} and \autoref{sec:entanglement} are conducted with five seeds.

In addition to sampling parameters randomly, we also train \glspl{qfm} on ten generated Fourier series (\cf~\autoref{sec:dataset}), while applying the aforementioned types of noise and circuits, using a noise level of 3\%.
Specifically, we observe as time series during the course the training,
\begin{enumerate*}[label=(\roman*)]
  \item the \gls{mse};
  \item the absolute coefficient values $\{\lvert c_{\bomega} \rvert\}$;
  \item the difference $\diffw$ to the target coefficients and
  \item the entangling capability
\end{enumerate*}

Our results are fully reproducible~\cite{Mauerer:2022}, with the code available in Ref.~\cite{franz_reproduction_2025}.

\subsection{Coefficients}
\label{sec:coefficients_exp}

We consider the effect of noise on the Fourier coefficients of a \gls{qfm}.
As the standard deviation across seeds is small (\ie, $<10\%$ for $\mu_c(\bomega)$), figures show the mean of $\mu_c(\bomega)$, $\sigma_c(\bomega)$ or $\Cov_c$ over the five seeds.
Samples for a coefficient with a mean $\mu_c(\bomega)$ below \num{1e-14} are taken to be zero, as the corresponding frequency $\bomega$ is effectively not contained in the spectrum.

\subsubsection{Input Encoding}

\begin{figure}[htbp] 
  \begin{subfigure}[c]{\subfigurewidth}
    \includegraphics[width=\columnwidth]{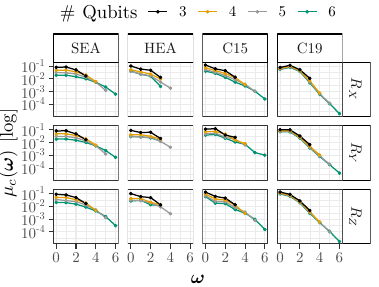}
    \caption{Absolute coefficient mean.}
    \label{fig:coefficients_mean_encoding}
  \end{subfigure}\hfill
  \begin{subfigure}[c]{\subfigurewidth}
    \includegraphics[width=\columnwidth]{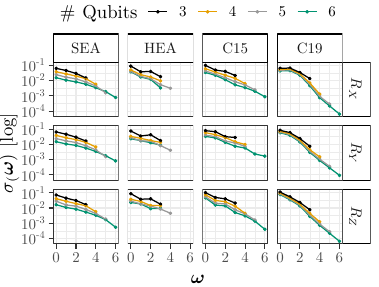}
    \caption{Relative standard deviation.}
    \label{fig:coefficients_sd_encoding}
  \end{subfigure}
  \caption{Absolute coefficient mean $\mu_c(\bomega)$ and the corresponding relative standard deviation $\sigma_c(\bomega)$ for $[3\dots 6]$ qubits and $R_{\{X, Y, Z\}}$ encodings over frequencies.}
  \label{fig:coefficients_encoding}
\end{figure}

As the encoding strategy substantially impacts coefficient values~\cite{caro_encoding-dependent_2021}, we first investigate in three different rotational Pauli encodings around $X$, $Y$, and $Z$ axis to encode one-dimensional data ($D = 1$) in a noiseless setting and observe how this not only changes the spectrum, but also the real and imaginary part.
\autoref{fig:coefficients_encoding} shows that all circuits suffer from a strong decay in the absolute coefficient value as well as in the corresponding relative standard deviation with increasing frequency. The magnitude of this decay varies with circuit structure, that is \gls{c19} exhibits lower $\mu_c(\bomega)$ at higher frequencies than the other \ase.
These findings are in line with Ref.~\cite{mhiri_constrained_2024}.
Generally speaking, the \as determines the spectral properties of the \gls{qfm} with $c_{\bomega}(\btheta)$ where the exact correlation between $\btheta$ and $\bomega$ varies.
Furthermore, the \gls{hea} with four to six qubits, and \gls{c15} with five qubits, resulted in $\mu_c(\bomega)$ smaller than \num{1e-14} for some frequencies, not exhibiting a full spectrum.
However, we note that the absolute coefficient for individual parameter samples is above the threshold, indicating that a full spectrum is possible with these \ase, but unlikely due to the sampling resolution in parameter space.

\begin{figure}[tbp]
  \begin{minipage}[c]{0.6\textwidth}
    \includegraphics[width=\columnwidth]{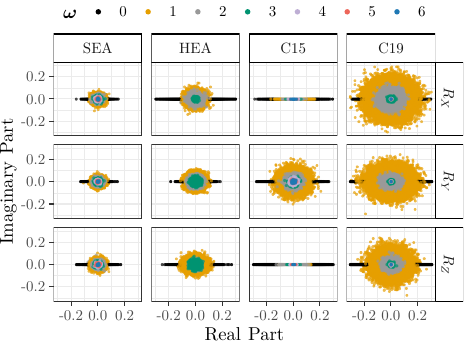}
  \end{minipage}\hfill
  \begin{minipage}[c]{0.38\textwidth}
    \caption{Coefficients, separated into real and imaginary parts for a circuit with six qubits and different single qubit Pauli-encodings. The individual frequency components are colour-coded.}
    \label{fig:coefficients_real_imag_encoding}
  \end{minipage}
\end{figure}

While there is no obvious difference in $\mu_c(\bomega)$ throughout the different encoding strategies, things change when we look at the real and imaginary parts of the individual coefficients $c_{\bomega}(\btheta)$ separately, as depicted in~\autoref{fig:coefficients_real_imag_encoding}.
Here, \gls{c15} exhibits an imaginary part in the coefficients, only for the case of an $R_Y$ encoding, missing a degree of freedom in the imaginary part for the $R_X$ and $R_Z$ encoding.
Therefore, in subsequent experiments on one-dimensional inputs, we use the $R_Y$ encoding for the results presented in the main text.

\subsubsection{Impact of Noise on the Real- and Imaginary Parts}
\label{sec:exp_covar}

\autoref{fig:coefficients_real_imag_encoding} suggests that coefficients are evenly distributed in the real and imaginary part, with no clear correlation (apart from \gls{c15} with $R_X$ and $R_Y$ encoding).
To corroborate this observation, we next compute the elements of the coefficients covariance matrix of~\autoref{eq:coeff_covar}.
We calculate coefficients for noiseless samples and for samples subjected to each noise type at 3\% probability.
Since all circuit sizes yield consistent results, we focus on six-qubit circuits and only on frequencies $\bomega \in \{0, 1, \text{max}\}$, where $\bomega = \text{max}$ refers to the highest occurring frequency in the noiseless case.

\begin{figure}[tbp]
  \begin{minipage}[c]{0.6\textwidth}
    \includegraphics[width=\columnwidth]{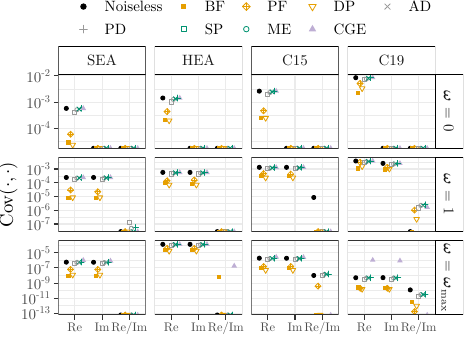}
  \end{minipage}\hfill
  \begin{minipage}[c]{0.38\textwidth}
    \caption{Elements of the coefficients covariance matrix in noiseless and noisy (3\% of each noise type) settings for six qubits. Points on the bottom border are zero. $\bomega = \bomega_\text{max}$ denotes the maximum frequency in the noiseless spectrum, that is $\bomega = 4$ for the \gls{hea} and $\bomega = 6$ for the remaining \ase.}
    \label{fig:coefficients_covar}
  \end{minipage}
\end{figure}
The results shown in~\autoref{fig:coefficients_covar}, align with those in~\autoref{fig:coefficients_real_imag_encoding}, demonstrating that the variances of both the real and imaginary parts of the coefficients are approximately equally high across configurations.
However, the covariance between these components is significantly smaller or even zero.
This suggests that the real and imaginary parts of a coefficient are not strongly correlated, allowing us to consider only the absolute value of the coefficients in subsequent experiments.
While additional variance experiments are detailed in~\autoref{sec:coeff_abs_exp},~\autoref{fig:coefficients_covar} highlights that decoherent gate errors significantly reduce the variance of the coefficients, regardless of circuit structure.
In contrast, other noise types appear to have no substantial impact on variance.

\begin{figure}[tbp]
  \begin{minipage}[c]{0.6\textwidth}
    \includegraphics[width=\columnwidth]{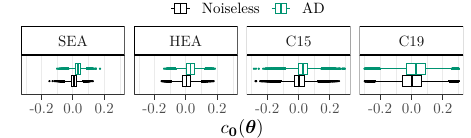}
  \end{minipage}\hfill
  \begin{minipage}[c]{0.38\textwidth}
    \caption{Real coefficient for $\bomega = \boldsymbol{0}$ in a noiseless setting vs. applying 3\% of \gls{ad} for circuits with six qubits.}
    \label{fig:coeffs_real_ad}
  \end{minipage}
\end{figure}

As shown in~\autoref{fig:coefficients_real_imag_encoding} coefficients are centred around zero for all configurations in a noiseless setting.
Generally, this also applies to noisy scenarios with the exception of the \gls{ad} channel, which shifts the coefficients of $\bomega = \boldsymbol{0}$ towards positive values, as shown in~\autoref{fig:coeffs_real_ad}.
This indicates that the offset due to the $Z$-basis measurement is skewed when applying an \gls{ad} channel.

\subsubsection{Impact of Noise on the Absolute Value}
\label{sec:coeff_abs_exp}

\begin{figure}[t]
  \begin{subfigure}[c]{\subfigurewidth}
    \includegraphics[width=\textwidth]{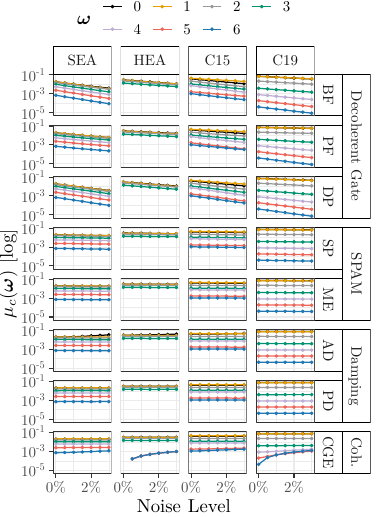}
    \caption{Absolute coefficient mean.}
    \label{fig:coefficients_mean}
  \end{subfigure}\hfill
  \begin{subfigure}[c]{\subfigurewidth}
    \includegraphics[width=\textwidth]{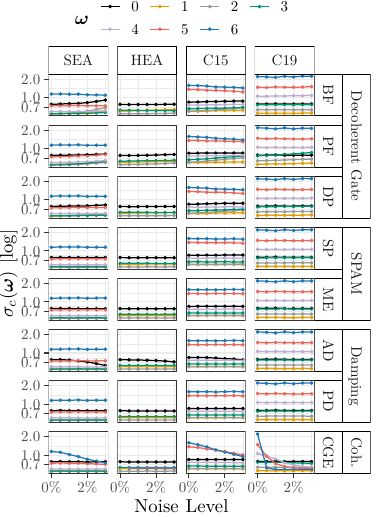}
    \caption{Relative standard deviation.}
    \label{fig:coefficients_var}
  \end{subfigure}
  \caption{
    Absolute coefficient mean $\mu_c(\bomega)$ and the corresponding relative standard deviation $\sigma_c(\bomega)$ over noise levels for various types of noise and six-qubit circuits that use $R_Y$ encoding.}
  \label{fig:coefficients_mean_var}
\end{figure}

Next, we examine how noise affects the coefficient mean and standard deviation across different \ase for one-dimensional inputs ($D=1$) using the $R_Y$ encoding.
The results are shown in~\autoref{fig:coefficients_mean_var} for each type of noise, evaluated on six-qubit circuits.
Additionally, we computed the coefficient values with an identical setup for the $R_X$ encoding, which we summarise in \autoref{apx:rx_results}.
These results demonstrate that there are many similarities, yet also fundamental differences for the coefficients using $R_X$ and $R_Y$ encodings, highlighting the importance of the encoding strategy selection.
Furthermore, we conducted the same set of experiments for circuits with three to five qubits, which provided similar outcomes throughout encodings, but accordingly with fewer frequencies.
A selection of the results is provided in~\autoref{apx:quant_exp} while the supplementary material in Ref.~\cite{franz_reproduction_2025} provides full result sets.

\autoref{fig:coefficients_mean} shows that decoherent gate errors generally lead to an exponential decay of coefficient mean for all considered \ase,
which is in line with the findings for uniform coefficient contraction under noise made in Ref.~\cite{fontana_classical_2023}.
For the \gls{spam} and damping noise the contraction is weaker, or not observable at all for $\mu_c(\bomega)$.

The relative standard deviation in~\autoref{fig:coefficients_var} remains constant for all noise types except \glspl{cge} for all frequencies and \ase, while frequencies $\bomega \neq \boldsymbol{0}$ generally lead to higher $\sigma_c(\bomega)$.
Since $\sigma_c(\bomega)$ is proportional to $\mu_c(\bomega)$, this implies that the absolute variance also decreases exponentially with increasing noise level under the influence of decoherent gate errors.
Interestingly, in case of the \gls{sea} with \gls{bf} applied, $\sigma_c(\bomega)$ for $\bomega = \boldsymbol{0}$ increases with increasing noise level, which likely is also a consequence of it being the relative standard deviation to the corresponding $\mu_c(\bomega)$, which in comparison only slightly increases.

Generally \gls{spam} and damping noise have minimal effect on both, coefficient mean and relative standard deviation.
While the influence of \gls{cge} is similar on the lower frequency coefficients, we observe that for the two highest frequencies $\mu_c(\bomega)$ increases, while $\sigma_c(\bomega)$ rapidly decreases with increasing noise level, especially for the \gls{hea} and \gls{c19} (see~\autoref{sec:coherent_exp} for a detailed discussion).

Given substantial similarities across different noise types and frequencies, we present a curated selection of combined results for $D \in \{1,2\}$ across all qubit numbers and \ase in~\autoref{apx:quant_exp}.
The comprehensive results are available in our supplementary material in Ref.~\cite{franz_reproduction_2025}.

\subsubsection{Effect of Coherent Noise}
\label{sec:coherent_exp}

\begin{figure}[t]
  \includegraphics[width=\columnwidth]{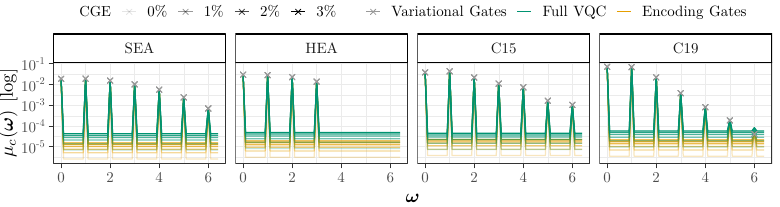}
  \caption{Absolute coefficient mean $\mu_c(\bomega)$ for 6 qubits and $R_Y$ encodings over frequencies for different noise levels $[0,\dots,3]\%$ and different \ase. Noise is applied either on the encoding gates ($\bepsilon_{\bx}$), the trainable gates ($\bepsilon_{\btheta}$) or on the full \gls{vqc} ($\bepsilon_{\bx}$ and $\bepsilon_{\btheta}$).}
  \label{fig:subsampling}
\end{figure}

As shown in \autoref{eq:cge}, the \gls{cge} on the trainable gates  $\bepsilon_{\btheta}$ only affects the magnitude of each $c_{\bomega}$.
However, the \gls{cge} on the encoding gates $\bepsilon_{\bx}$ alters the frequency component as the differences in the eigenvalues $\Lambda$ do not fully overlap, yielding a higher number of unique frequency components while the redundancies $R(\omega)$ are reduced.
This effect becomes visible when increasing the frequency resolution $\Delta f$ by increasing the window length of the \gls{fft}.
In \autoref{fig:subsampling} we can observe the effect of \gls{cge} on the spectrum of different 6-qubit \ase with a window length being 10 times the required minimum according to the sampling theorem.
Here, we differentiate between applying the \gls{cge} on
\begin{enumerate*}[label=(\roman*)]
  \item the encoding gates ($\bepsilon_{\bx}$);
  \item the trainable gates ($\bepsilon_{\btheta}$) or
  \item on the full \gls{vqc} ($\bepsilon_{\bx}$ and $\bepsilon_{\btheta}$).
\end{enumerate*}

While the absolute mean value of the integer-valued frequencies $[0,1,\dots,6]$ remains mostly constant for all noise levels, we can observe that the magnitude of non-integer frequencies in between increases.
Notably, only the input \gls{cge} causes this effect, which is amplified in combination with but not solely caused by the \gls{cge} acting on the variational gates.

This observation is in line with an approach introduced in~\cite{jaderberg_let_2024}, where trainable parameters were added to the input encoding, increasing the number of differences between the summation of eigenvalues $\Lambda$ and therefore also increasing the number of frequencies.
While the referenced work used this effect to adjust the frequencies individually, the \gls{cge} can be viewed similarly while changing the frequencies randomly and therefore, on average, causing a uniform increase of non-integer frequencies.

\begin{figure}[tbp]
  \begin{minipage}[c]{0.6\textwidth}
    \includegraphics[width=\columnwidth]{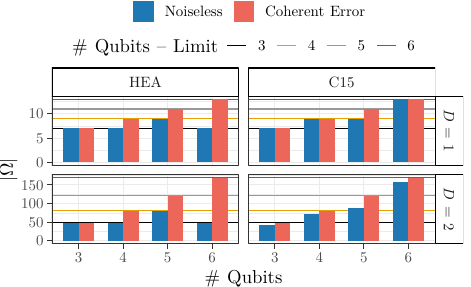}
  \end{minipage}\hfill
  \begin{minipage}[c]{0.38\textwidth}
    \caption{Number of frequencies in the spectrum with and without applying a \gls{cge} for $D$-dimensional inputs and only those circuits where the maximum possible number of frequencies is not achieved in the noiseless case.}
    \label{fig:n_freqs_coherent}
  \end{minipage}
\end{figure}

The impact of \gls{cge} on the frequency spectrum also becomes evident when comparing the number of frequencies with and without coherent noise, as illustrated in~\autoref{fig:n_freqs_coherent}.
In particular, coefficients that are computed to be zero in the \gls{fft} under noiseless conditions become non-zero in the presence of \gls{cge}.
Therefore, the increase in higher-frequency coefficients, as shown in \autoref{fig:coefficients_mean_var} and \autoref{fig:coefficients_mean_1D_2D}, is a result of these spectral shifts, which lead to a more uniform distribution of coefficients.
This also applies to the \gls{hea} and \gls{c15} that do not result in a complete frequency spectrum, likely due to limitations in the \as structure and input encodings, which restrict the set of achievable eigenvalue differences (\cf~\autoref{eq:set_of_frequencies}).
While \glspl{cge} seems to solve this problem, we would like to point out that although the circuit under \gls{cge} contains more frequencies, this does not mean that they are individually tunable (\cf \autoref{sec:training_exp}).
This can result in a spectrum that is significantly correlated, a phenomenon that is undesirable in the context of most learning problems.

\subsubsection{Coefficients during Training}
\label{sec:training_exp}

\begin{figure}[b]
  \begin{minipage}[c]{0.6\textwidth}
    \includegraphics[width=\columnwidth]{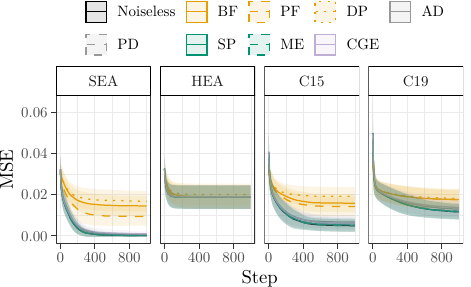}
  \end{minipage}\hfill
  \begin{minipage}[c]{0.38\textwidth}
    \caption{The \gls{mse} over ten parameter initialisation seeds and ten randomly generated problem instances during training. Lines represent the mean, and shaded areas show the standard deviation over the $10 \times 10$ configurations.}
    \label{fig:training_mse}
  \end{minipage}
\end{figure}

To measure the training performance under noisy conditions, we utilise the setup and metrics described in \autoref{sec:dataset} for the synthetic regression tasks.
Training is performed using the Adam optimiser~\cite{kingma2017adammethodstochasticoptimization} with a learning rate of \num{0.01} over \num{1000} training steps.
The \gls{mse} over the course of the training, shown in \autoref{fig:training_mse}, demonstrates that all configurations converge although the performance between circuits differs:
The \gls{hea} and \gls{c19} fail to converge towards the optimum for all tested problem- and model initialisations, while \gls{c15} and the \gls{sea} achieve lower \glspl{mse}, with the \gls{sea} performing best on this metric in a noiseless setting.

The detrimental effect of noise is particularly evident for the decoherent gate errors across all \ase, resulting in higher \glspl{mse}, while the other noise types appear to be less significant.
It is evident that the distance in the \gls{mse} between the noiseless case and the \glspl{qfm} that suffer from decoherent gate errors increases as the \gls{mse} in the noiseless case decreases.
This results in a similar \gls{mse} of $\approx 0.02$, when decoherent gate errors are applied throughout all \ase.
Other types of noise result in minor deviations from the noiseless case.

\begin{figure}[tbp]
  \includegraphics[width=\columnwidth]{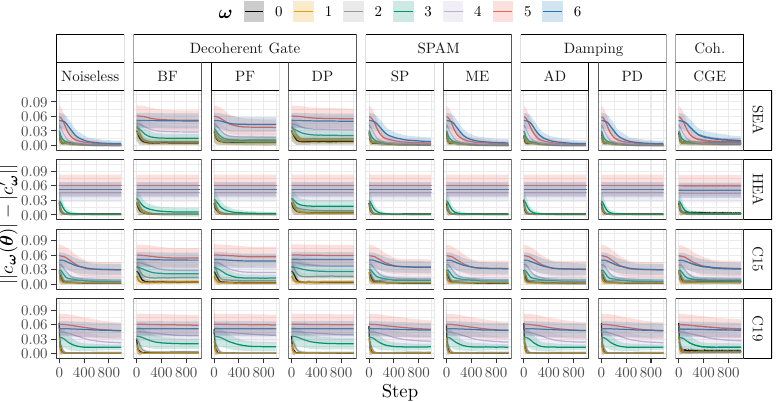}
  \caption{The difference $\diffw$ between the target and learned coefficients $c_{\bomega(\btheta)}$ and $c'_{\bomega}$ respectively of the \glspl{qfm} averaged over ten model parameter initialisation seeds and ten randomly generated problem instances during training. Lines represent the mean, and shaded areas show the standard deviation over the $10 \times 10$ configurations.}
  \label{fig:training_coefficient_distance}
\end{figure}

The trend of the \gls{mse} can be attributed to how closely the coefficients of the \gls{qfm} approach the target coefficients.
In \autoref{apx:training_instances}, we show the exact coefficients for one of the problem instances considered over the course of the training where we note, that noise causes a shift of the frequencies as they are affected differently by noise.
\autoref{fig:training_coefficient_distance} shows the mean absolute coefficient difference $\diffw$ between the \glspl{qfm} and the objective Fourier series over all problem seeds.
Considering the noiseless case, while $\diffw$ for lower $\bomega$ converges faster to zero across all \ase, the \ase, which achieve a lower \gls{mse}, also reach a lower $\diffw$ for all $\bomega$.
Nevertheless, higher frequencies are also crucial for the performance, as indicated by the \gls{hea}, which achieves a low $\diffw$ for $\bomega \in [0\dots4]$, however with a constant high $\diffw$ for the remaining two highest frequencies, as it does not achieve a full spectrum, resulting in a overall higher \gls{mse}.
The large values of $\diffw$ for higher frequencies across all \ase, except the \gls{sea}, indicate that the Fourier coefficients are not fully tunable for every \as, even in a noiseless setting.

When adding noise, there are only minor differences between the performance for the noiseless setting, and \gls{spam}, Damping and \glspl{cge}.
For the \gls{cge}, we notice small variations in the coefficient values between training steps, evident in \autoref{fig:training_seed1000}, while the remaining noise types lead to more smooth curves.
While these variations also occur for the higher frequencies of the \gls{hea} that do not appear in the noiseless spectrum, they do not contribute to optimising the coefficients towards the target.
Decoherent gate errors lead to more drastic performance issues, visible in both, the \gls{mse} and $\diffw$.

\subsection{Expressibility}

\begin{figure}[tbp]
  \includegraphics[width=\columnwidth]{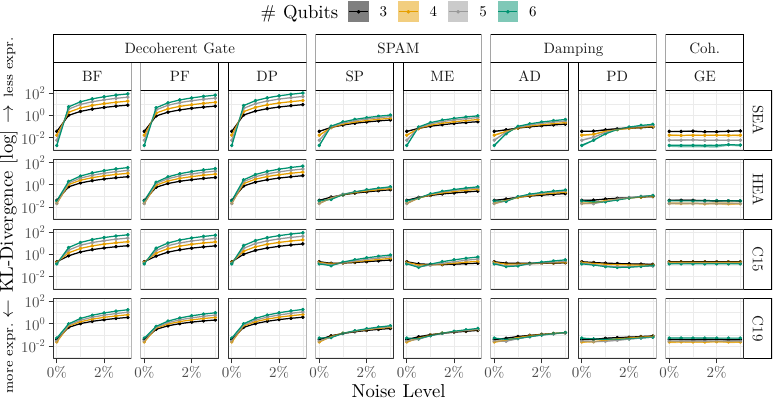}
  \caption{Average Expressibility (\ie, inverse of the \gls{kl} divergence) over randomly sampled parameters under the influence of increasing noise levels. The points represent the mean and the shaded areas around it refer to the minimum and maximum across all five seeds, which is negligible in most cases as we chose a sufficiently large number of samples.}
  \label{fig:expressibility}
\end{figure}

The expressibility is calculated as introduced in~\autoref{sec:expressibility} and quantified using the \gls{kl}-divergence to the Haar distribution (\cf~\autoref{eq:kl_divergence}).
The experimental results for all \ase and qubit counts are presented in~\autoref{fig:expressibility}.

It can be observed that all decoherent gate errors lead to an increase in the \gls{kl}-divergence, thereby reducing expressibility.
This effect is also consistent across \gls{spam} and damping noise, though the impact is less pronounced.
The observed decrease in expressibility aligns with the reduced variance of coefficients noted in~\autoref{sec:coefficients_exp} and the connection to frequency redundancies discussed in Ref.~\cite{mhiri_constrained_2024}.

Coherent noise, on the other hand, has no measurable effect on expressibility.
This is not surprising since we omit the encoding gates for the \gls{cge} to act upon, leaving the system in a coherent state.
Effects of shifts in the trainable parameters appear to cancel out over the parameter samples.

In a noiseless environment, the \gls{sea} achieves a higher expressibility compared to the other \ase.
However, when coherent noise is applied, the expressibility is quickly equalised.
While the effect is consistent across all circuits tested, it can be observed that it becomes more pronounced as the number of qubits increases.
This suggests that all forms of decoherent noise degrade the quantum nature of the circuits.

\subsection{Entanglement}
\label{sec:entanglement}

We first utilise the \gls{mw} measure as introduced in~\autoref{sec:entangling_capability_mw} to compute the entanglement of the circuits without noise.
Subsequently, the \gls{ef} (\cf~\autoref{sec:entangling_capability_ef}) is used to measure the entangling capability for the mixed states with increasing noise levels.

\begin{figure}[tbp]
  \begin{minipage}[c]{0.5\textwidth}
    \includegraphics[width=\columnwidth]{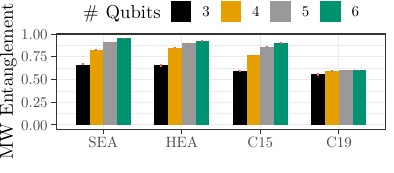}
  \end{minipage}\hfill
  \begin{minipage}[c]{0.48\textwidth}
    \caption{Average \gls{mw} entanglement over randomly sampled parameters for the considered \ase in a noiseless environment.
      The bars represent the mean values with error bars from the minimum to maximum highlighted in red across the five seeds.
    }
    \label{fig:ent_mw}
  \end{minipage}
\end{figure}

The results for the noiseless circuits are shown in \autoref{fig:ent_mw}.
Here, it can be observed, that an increasing qubit count leads to a higher overall value for the entangling capability, while the specific values for the \gls{sea}, \gls{hea} and \gls{c15} are similar.
\gls{c19}, which has controlled rotation gates as entangling gates, results in a lower entangling capability for more than 4 qubits.

\begin{figure}[tbp]
  \includegraphics[width=\columnwidth]{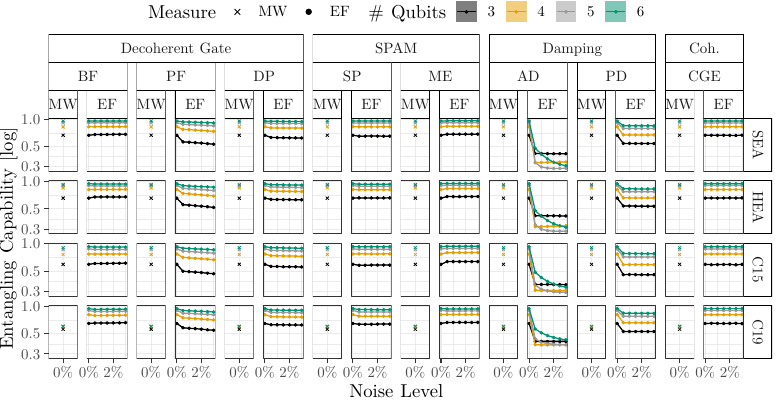}
  \caption{Average \gls{mw} entangling capability and \gls{ef} over randomly sampled parameters under the influence of increasing noise levels.
    Points represent the mean of five seeds and lines are a linear interpolation to guide the eye.
    Shaded areas represent the maximum/minimum entanglement across the five seeds.
    As the difference between seeds is small, it is not visible in the plot.
  }
  \label{fig:entanglement}
\end{figure}

The results for the mixed states using the \gls{ef} measure are shown in \autoref{fig:entanglement}.
Notably, \gls{ef} results in different values than the \gls{mw} measure, also for noiseless circuits, especially observable for \gls{c19}.
This observation may be a consequence of the properties of the \gls{ef} measure:
At zero noise-probability for the \gls{ef}, an eigendecomposition results in only one pure state, over which the \gls{mw} measure is computed.
However, this decomposition is not unique (\cf~\autoref{sec:entangling_capability}), and only one of many possibilities, potentially resulting in different values for the entangling capability.
Therefore, we interpret the values in~\autoref{fig:entanglement} as upper bounds of the entangling capability, which drop below the \gls{mw} entangling capability with applied noise in most cases.

Within the decoherent gate-, and \gls{spam}-error groups a similar behaviour can be observed for each \as.
The values here slightly decrease or remain constant for both decoherent gate- and \gls{spam}-errors, with some exceptions for three and four qubits with \gls{bf}- and \gls{me}-noise, where the entanglement slightly increases from 0\%-level noise to 0.5\%.
Similarly, for \gls{pd}, the entangling capability also decreases with noise for all \ase.
For \gls{ad}-noise entangling capability decreases more drastically for all circuits, and number of qubits with increasing noise level.
\glspl{cge} do not have an observable impact on the \gls{ef}.
Our results suggest that noise is overall detrimental for the entangling capability, especially for \gls{ad}-noise.

As we discuss in \autoref{apx:entanglement_training}, the entangling capability of a \gls{qfm} decreases over the course of a training for the considered learning task similarly across all \ase, indicating no clear correlation between training performance and entangling capability.
When adding noise, we measure a similar impact on the overall entangling capability during training as in \autoref{fig:entanglement}, which is slightly lower than in the noiseless case for most \ase.

In the broad and systematic empirical analysis presented in this section, we find that the detailed effect of different types of noise leads to the main general findings for \gls{qml} outlined as follows:
\begin{tcolorbox}[width=\textwidth, sharp corners, colback=lfd4!10, colframe=lfd4, title=Result Summary]
  The influence of noise on properties of \glspl{qfm}, particularly the Fourier spectrum, expressibility and entangling capability, can be predicted.
  The effects of each noise type are uniform throughout all tested circuits.
  More importantly, the structure of a \gls{vqc} and the input encoding have a crucial impact on these properties, independently of the types of noise applied, also in a noiseless setting.
  This is a relevant finding for the mid- and long-term of prospective \glspl{ftqc}.
\end{tcolorbox}

\section{Discussion}
\label{sec:discussion}

This work explores the effects that noise has on the coefficients, expressibility and entangling capability of \glspl{qfm}.
In particular,
\begin{enumerate*}[label=(\roman*)]
  \item our work confirms the anti-correlation between coefficients and noise observed in Ref.~\cite{fontana_classical_2023}, and demonstrates that the decrease of coefficient values is even exponential for decoherent gate errors.
        Furthermore, we observed that all of the tested \ase are less susceptible to damping or \gls{spam} errors.
  \item Using the example of learning a random Fourier series, we demonstrate that decoherent errors have a detrimental influence on the capabilities of \glspl{qfm} to assume certain Fourier coefficients.
        This directly impacts the kinds of functions that can be approximated by a \gls{qfm}.
  \item Our experiments show that the expressibility, measured by the distance to the Haar distribution and the coefficients variance also vanishes (potentially exponentially) with the absolute coefficients value with increasing noise level.
  \item The entangling capability, which is a general indicator of the \emph{quantumness} of a \gls{vqc}, also decreases when subject to noise.
\end{enumerate*}

The \ase investigated in this work show similar patterns for all of the above.
Although our findings are based on a limited circuit size and number of input dimensions (up to six qubits and two features), they demonstrate consistent patterns as the number of qubits increases from three to six.
This suggest potential for broader applicability in larger circuits.
Additionally, similar results across one-dimensional and two-dimensional inputs offer a good indication for generalisation despite the exponential scaling of frequencies in the spectrum.

Although the observation of the detrimental influence of noise may not be a novel finding~\cite{fontana_spectral_2022, Dalzell_2024}, to the best of our knowledge our work is the first to show this effect in a holistic, systematic manner on the fundamental computational aspects of \glspl{vqc}.

As each learning problem in \gls{qml} has different requirements on the \as that is employed, a general \emph{all-fits-one} \as is unlikely to exist.
Nevertheless, based on our experiments, we can make some statements on the measures of quality for the tested \ase:
\begin{enumerate*}[label=(\roman*)]
  \item The \textbf{\gls{sea}} is, as the name suggests the \as with the highest entangling capability, although the difference to \gls{hea} and \gls{c15} in that regard is small.
        Among the tested \ase, the \gls{sea} also utilises most trainable parameters per qubit, resulting in a comparatively higher expressibility in a noiseless setting (\cf~\autoref{fig:expressibility}), which is beneficial for the considered learning tasks.
        However, the expressibility quickly vanishes when applying noise.
        The corresponding Fourier spectrum is full (at least for up to two input dimensions), and comparatively uniformly distributed (\cf~\autoref{fig:coefficients_mean_encoding}).
  \item In the \textbf{\gls{hea}}, the amount by which the depth of the entangling sequence grows, is constant, compared to the other \ase, where it scales linearly with the number of qubits.
        This desirable property comes with the cost of lacking the full Fourier spectrum on average, even at one input dimension, which directly impacts the kinds of functions that can be learned by the \gls{hea}.
  \item \textbf{\gls{c15}} demonstrates similar behaviour to the \gls{sea} and \gls{hea} in the scaling of entanglement and expressibility under noise.
        It further has a unique characteristic, where the imaginary part in the coefficients appears exclusively for a $R_Y$ encoding.
        As the \gls{hea}, \gls{c15} does not consistently achieve a full spectrum.
  \item \textbf{\gls{c19}} achieves a full spectrum for up to two input dimensions, as the \gls{sea}.
        However, the coefficients vanish quickly with increasing frequency, even in the noiseless case (\cf~\autoref{fig:coefficients_mean_encoding}), indicating that not all parts of the spectrum may be effectively utilised.
        Despite having the lowest \gls{mw} entangling capability, \gls{c19} features tunable parameters in the controlled-$R_X$ gates, which could potentially allow for adjusting this property.
\end{enumerate*}

\section{Conclusion and Outlook}
\label{sec:conclusion}

Given the potential for conducting an extensive array of numerical experiments on this subject, the results presented are expected to offer researchers valuable insights into the behaviour of \gls{qfm} under noisy conditions.
However, an analytical correlation between these factors and potential generalisations to larger circuit sizes (both in the number of qubits and the circuit depth) remains to be investigated in future research.
Also how different \ase or encoding strategies, such as the ones presented in Ref.~\cite{casas_multidimensional_2023} fit into this pattern remains to be explored.

While we acknowledge that noise will lose its importance once \gls{ftqc} is established, the road to \gls{ftqc} remains stony, and as not every hardware architecture is suitable for error correction, the effects of noise on \gls{qfm} are still relevant as the proposed results are hardware agnostic.
In subsequent studies, we intend to perform analytical derivations to further extend the results presented and allow more definitive conclusions to be drawn, also for arbitrary \ase.

\vspace{0.2em}
\noindent\textbf{Acknowledgements}
MS, LC, EK and AS acknowledge support by the state of Baden-W\"urttemberg through bwHPC.
MF and WM acknowledge support by the German Federal Ministry of Research, Technology and Space (BMFTR), funding program \enquote{Quantum Technologies—--From Basic Research to Market}, grant number 13N16092.
WM acknowledges support by the High-Tech Agenda of the Free State of Bavaria.

\printbibliography

\clearpage
\appendix

\section{Noise Models}
\label{apx:noise}

In the following, we describe the utilised methods to model both, decoherent and coherent errors.
For the decoherent errors we utilise the established Kraus formalism that models the
(possibly non-unitary) effects of noise on a density
operator \(\rho\) by
\(\rho \mapsto \sum_i K_i \rho K_i^\dagger\) with Kraus operators
\(\{K_{i}\}\).

\subsection{Decoherent Gate Errors}
We model decoherent gate errors by a combination of
Pauli operations $\{I, X, Y, Z\}$ applied with a certain probability after each (ideal and noiseless) quantum gate.
For simplicity, we only consider one-qubit channels, summarised in \autoref{tab:kraus}. 

\begin{table}[b]
  \centering
  \begin{tabular}{l|l}
    \toprule
    \textbf{Noise Type}      & \textbf{Kraus Operators}      \\
    \midrule
    \textbf{Decoherent Gate} &                               \\[0.7em]
    \glsxtrfull{bf}          &
    \(
    K_{0}=\sqrt{1-p_\text{bf}}I, \quad
    K_{1}=\sqrt{p_\text{bf}}X
    \)
    \\[0.7em]
    \glsxtrfull{pf}          &
    \(
    K_{0}=\sqrt{1-p_\text{pf}}I, \quad
    K_{1}=\sqrt{p_\text{pf}}Z
    \)                                                       \\[0.7em]
    \glsxtrfull{dp}          &
    \(\begin{aligned}
        K_{0} =\sqrt{1-p_\text{dp}}I, \quad
         & K_{1} =\sqrt{p_\text{dp} / 3}X, \\
        K_{2} =\sqrt{p_\text{dp} / 3}Y, \quad
         & K_{3} =\sqrt{p_\text{dp} / 3}Z
      \end{aligned}\) \\
    \midrule
    \textbf{Damping}         &                               \\[0.7em]
    \glsxtrfull{ad}          &
    \(
    K_{0}  =\left[\begin{smallmatrix}1 & 0 \\ 0 & \sqrt{1-p_\text{ad}}\end{smallmatrix}\right], \quad
    K_{1}  =\left[\begin{smallmatrix}0 & \sqrt{p_\text{ad}} \\ 0 & 0\end{smallmatrix}\right]
    \)                                                       \\[0.7em]
    \glsxtrfull{pd}          &
    \(
    K_{0} =\left[\begin{smallmatrix}1 & 0 \\ 0 & \sqrt{1-p_\text{pd}}\end{smallmatrix}\right], \quad
    K_{1} =\left[\begin{smallmatrix}0 & 0 \\ 0 & \sqrt{p_\text{pd}}\end{smallmatrix}\right]
    \)                                                       \\
    \bottomrule
  \end{tabular}
  \caption{Kraus operators for the decoherent gate and damping errors with the probabilities $p_{\text{type}}$ for each noise type.}
  \label{tab:kraus}
\end{table}

\subsection{Damping Errors}
\emph{Damping} errors correspond to relaxation and dephasing effects;
the probabilities depend on the relaxation time $T_1$ and dephasing time $T_2$ of a given \gls{qpu} implementation (we apply the damping channels at the end of a circuit.).
As we do are not interested in vendor-specific details, we
simply assume fixed probabilities.

\glsxtrfull{ad} describes the natural decay of the excited state $\ket{1}$ to the ground state $\ket{0}$ due to energy exchange with the environment.
Similarly, \glsxtrfull{pd} describes the transition of a quantum system towards classical behaviour y loosing coherence. Corresponding Kraus operators are listed in \autoref{tab:kraus}.

\subsection{State Preparation and Measurement Errors}

\glsxtrfull{spam} on a real quantum device can be faulty, just like any other operation.
Ideally, the initial state is $\ket{0}^{\otimes n}$ state, and a measurement returns either 0 or 1.
However, there is a probability that \gls{sp} is faulty, or that measurements yield 1 instead of 0, and vice versa.
We model \gls{sp} errors with \gls{bf} applied at the beginning of a circuit with probability $p_\text{sp}$.
\glsxtrfull{me} errors are applied at the very end of a circuit with probability $p_\text{me}$, again using \gls{bf} errors.

\subsection{Coherent Gate Error}

Notable, since each operation is applied to the quantum system using an imperfect real \gls{qpu}, the actual gate operation may deviate from the intended one.

The \glspl{cge} are modelled by adding $\bepsilon_{\bx} \sim \mathcal{N}(0, p_\text{cge}^2)$ and $\bepsilon_{\btheta} \sim \mathcal{N}(0, p_\text{cge}^2)$ to any parameterised rotational input and trainable gate, respectively.
The resulting Fourier series therefore becomes
\begin{equation}
  \label{eq:cge}
  f_\text{cge}(\bx + \bepsilon, \btheta + \bepsilon_{\btheta}) = \sum_{\bomega \in \bOmega} c_{\bomega}(\btheta + \bepsilon_{\btheta}) e^{\I \bomega^T (\bx + \bepsilon_{\bx})}
  = \sum_{\bomega \in \bOmega} c_{\bomega}(\btheta + \bepsilon_{\btheta}) e^{\I (\bomega + \bepsilon_{\bx})^T \bx}.
\end{equation}

If not stated explicitly, we consider the gate error on the encoding gates constant when calculating the coefficients of the model.
The reason for this is, that we want to investigate the spectrum as a property of a model, which would not be possible if the gate error acts independently on the encoding gates, as it presumably causes a shift in the frequency components.

\section{Ansätze}
\label{apx:ansaetze}

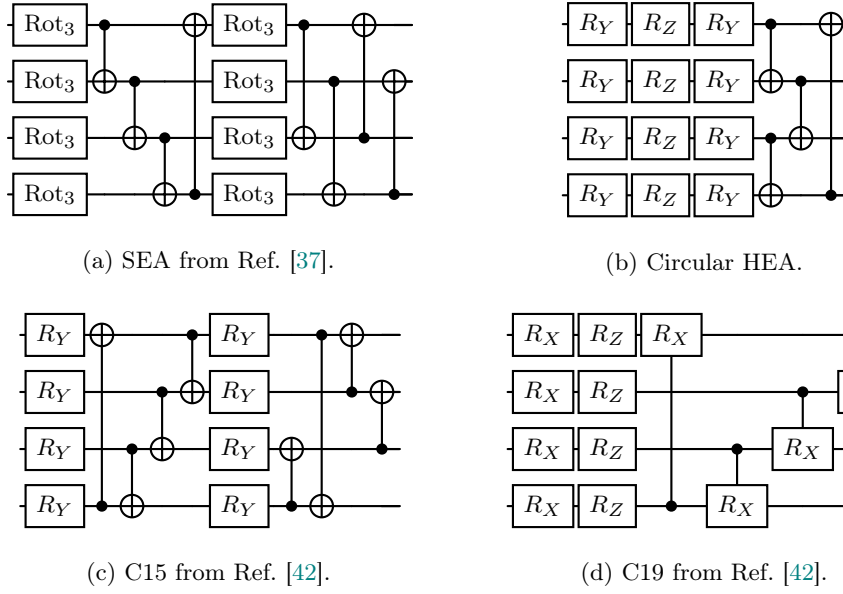
\begin{figure}[t]
  \newcommand{\Rot}{\text{Rot}}\centering
  \begin{subfigure}[c]{0.49\columnwidth}
    \centering
    \begin{tikzpicture}
      \node[] {
        \scriptsize\begin{quantikz}[row sep=0.2cm, column sep=0.7mm]
          \qw & \gate{\Rot_3} & \ctrl{1} &  &  & \targ{} & \gate{\Rot_3} & \ctrl{2} &  & \targ{} & &    \qw \\
          \qw & \gate{\Rot_3} & \targ{} & \ctrl{1} &  &  & \gate{\Rot_3} &  & \ctrl{2} &  & \targ{} &   \qw \\
          \qw & \gate{\Rot_3} &  & \targ{} & \ctrl{1} &  & \gate{\Rot_3} & \targ{} &  & \ctrl{-2} & &   \qw \\
          \qw & \gate{\Rot_3} &  &  & \targ{} & \ctrl{-3} & \gate{\Rot_3} &  & \targ{} &  & \ctrl{-2} & \qw
        \end{quantikz}
      };
    \end{tikzpicture}
    \caption{\gls{sea} from Ref.~\cite{schuld_circuit-centric_2020}.}
    \label{fig:strongly_entangling_ansatz}
  \end{subfigure}
  \begin{subfigure}[c]{0.49\columnwidth}
    \centering
    \begin{tikzpicture}
      \node[] {
        \scriptsize\begin{quantikz}[row sep=0.2cm, column sep=0.7mm]
          \qw & \gate{R_Y} & \gate{R_Z} & \gate{R_Y} & \ctrl{1} &  & \targ{}  & \qw \\
          \qw & \gate{R_Y} & \gate{R_Z} & \gate{R_Y} & \targ{} & \ctrl{1} &   & \qw \\
          \qw & \gate{R_Y} & \gate{R_Z} & \gate{R_Y} & \ctrl{1} & \targ{} &   & \qw \\
          \qw & \gate{R_Y} & \gate{R_Z} & \gate{R_Y} & \targ{} &  & \ctrl{-3} & \qw
        \end{quantikz}
      };
    \end{tikzpicture}
    \caption{Circular \gls{hea}.}
    \label{fig:hardware_efficient_ansatz}
  \end{subfigure}
  \begin{subfigure}[c]{0.49\columnwidth}
    \centering
    \begin{tikzpicture}
      \node[] {
        \scriptsize\begin{quantikz}[row sep=0.2cm, column sep=0.7mm]
          \qw & \gate{R_Y} & \targ{} &  &  & \ctrl{1} & \gate{R_Y} & & & \ctrl{3} & \targ{} &  & \qw \\
          \qw & \gate{R_Y} &  &  & \ctrl{1} & \targ{} & \gate{R_Y} & & &  & \ctrl{-1} & \targ{} & \qw \\
          \qw & \gate{R_Y} &  & \ctrl{1} & \targ{} & & \gate{R_Y} & & \targ{} &  &  & \ctrl{-1} & \qw \\
          \qw & \gate{R_Y} & \ctrl{-3} & \targ{} & & & \gate{R_Y} &  & \ctrl{-1} & \targ{} &  &  & \qw
        \end{quantikz}
      };
    \end{tikzpicture}
    \caption{\gls{c15} from Ref.~\cite{sim_expressibility_2019}.}
    \label{fig:circuit_15}
  \end{subfigure}
  \begin{subfigure}[c]{0.49\columnwidth}
    \centering
    \begin{tikzpicture}
      \node[] {
        \scriptsize\begin{quantikz}[row sep=0.2cm, column sep=0.7mm]
          \qw & \gate{R_X} & \gate{R_Z} & \gate{R_X} &  &  & \ctrl{1} & \qw \\
          \qw & \gate{R_X} & \gate{R_Z} &  &  & \ctrl{1} & \gate{R_X} & \qw \\
          \qw & \gate{R_X} & \gate{R_Z} &  & \ctrl{1} & \gate{R_X} &  & \qw \\
          \qw & \gate{R_X} & \gate{R_Z} & \ctrl{-3} & \gate{R_X} &  &  & \qw
        \end{quantikz}
      };
    \end{tikzpicture}
    \caption{\gls{c19} from Ref.~\cite{sim_expressibility_2019}.}
    \label{fig:circuit_19}
  \end{subfigure}
  \caption{\Ase investigated in this work, exemplarily for \num{4} qubits.
    $\Rot_3$ represents an arbitrary single qubit rotational gate, which takes three parameters.
    The single qubit Pauli-rotation gates $R_{\{X, Y, Z\}}$ are parametrised by one parameter each.}
  \label{fig:ansaetze}
\end{figure}

We use the \ase as depicted in \autoref{fig:ansaetze} in the experiments of our work.

\section{Coefficient Results for Pauli-RX Encoding}
\label{apx:rx_results}

\begin{figure}[t]
  \begin{subfigure}[c]{\subfigurewidth}
    \includegraphics[width=\textwidth]{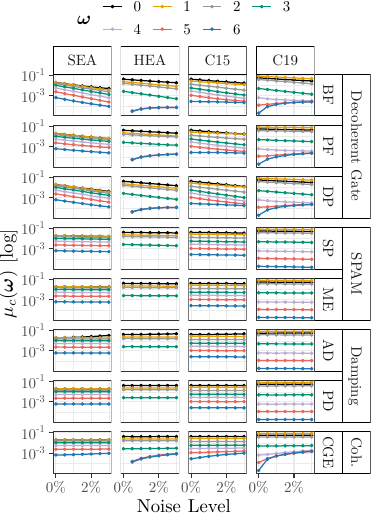}
    \caption{Absolute coefficient mean.}
    \label{fig:coefficients_mean_rx}
  \end{subfigure}\hfill
  \begin{subfigure}[c]{\subfigurewidth}
    \includegraphics[width=\textwidth]{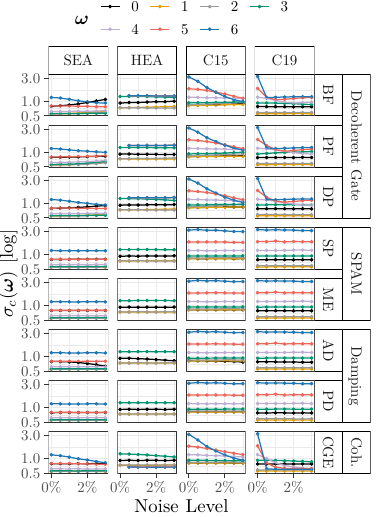}
    \caption{Relative standard deviation.}
    \label{fig:coefficients_var_rx}
  \end{subfigure}
  \caption{
    Analogue to \autoref{fig:coefficients_mean_var} but for $R_X$, instead of $R_Y$ encoding. Absolute coefficient mean $\mu_c(\bomega)$ and the corresponding relative standard deviation $\sigma_c(\bomega)$ over noise levels for various types of noise and six-qubit circuits that use $R_X$ encoding.}
  \label{fig:coefficients_mean_var_rx}
\end{figure}

In \autoref{fig:coefficients_mean_var_rx} we show the absolute coefficient mean $\mu_c(\bomega)$ and the corresponding standard deviation $\sigma_c(\bomega)$ for six-qubit circuits using $R_X$ encodings, analogous to \autoref{fig:coefficients_mean_var}.
While there are many similarities to the $R_Y$ encoding experiments, the main difference is that higher-frequency coefficients for the \gls{hea} and \gls{c19} increase when applying decoherent gate errors, while they decrease for the $R_Y$ encoding.
For the \gls{hea} this results in the higher frequencies becoming part of the spectrum, even in they are not present in the noiseless case, similar to when applying \glspl{cge}.
This is likely to random distortions in the expectation value, leading to an increase of all frequencies in the spectral representation that become more apparent at the low-valued coefficients at higher frequencies.
We also observe and discuss this effect in \autoref{apx:quant_exp}.

Additionally, \gls{c15} and \gls{c19} provide the highest standard deviation at 0\% noise level, especially for higher frequencies, which then decays (more rapidly for \gls{c19}) when decoherent or coherent gate errors are added.
For the remaining noise types and \ase $\sigma_c(\bomega)$ remains constant or slightly decreases (with the same exceptions for the \gls{sea} as in \autoref{fig:coefficients_var}).

\section{Quantitative Coefficient Results}
\label{apx:quant_exp}

To assess the impact of noise on circuits with multiple input features, we conduct the same set of experiments detailed in~\autoref{sec:coeff_abs_exp} and \autoref{apx:rx_results}, also for $D=2$, with the difference of only sampling over two instead of five random seeds for all qubit numbers to save computational resources.
While different encoding strategies are presented in Ref.~\cite{casas_multidimensional_2023}, this work focuses on a subsequent Pauli-encoding along different rotational axes (\ie, $X$ and $Y$ in our experiments).

\begin{figure}[tbp]
  \includegraphics[width=\columnwidth]{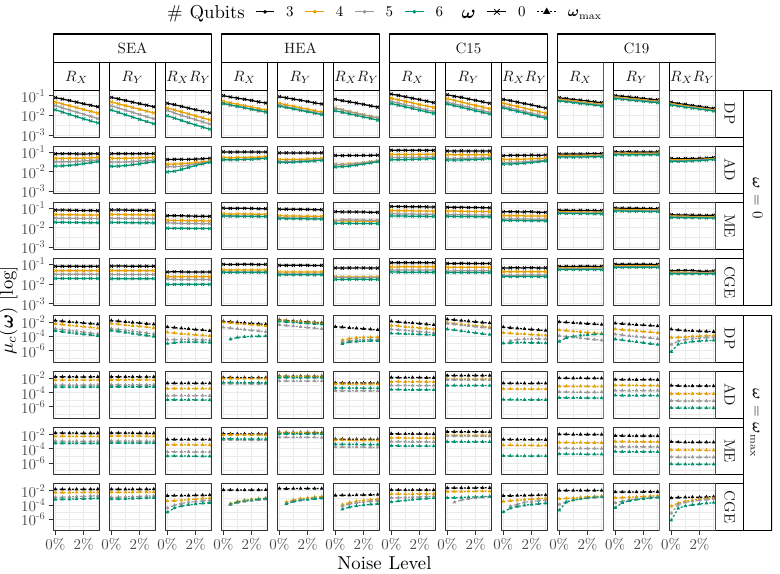}
  \caption{Absolute coefficient mean $\mu_c(\bomega)$ for the lowest frequency $\bomega=\boldsymbol{0}$ and highest frequency $\bomega = \bomega_\text{max}$ in the respective spectrum under the influence of varying noise levels. We considered the one-dimensional encodings $R_X$ and $R_Y$ ($D$ = 1), and a two-dimensional encoding $R_XR_Y$ ($D$ = 2). The y-axis for each facet row are equal throughout the respective $\bomega$, but differs in between.}
  \label{fig:coefficients_mean_1D_2D}
\end{figure}

\begin{figure}[tbp]
  \includegraphics[width=\columnwidth]{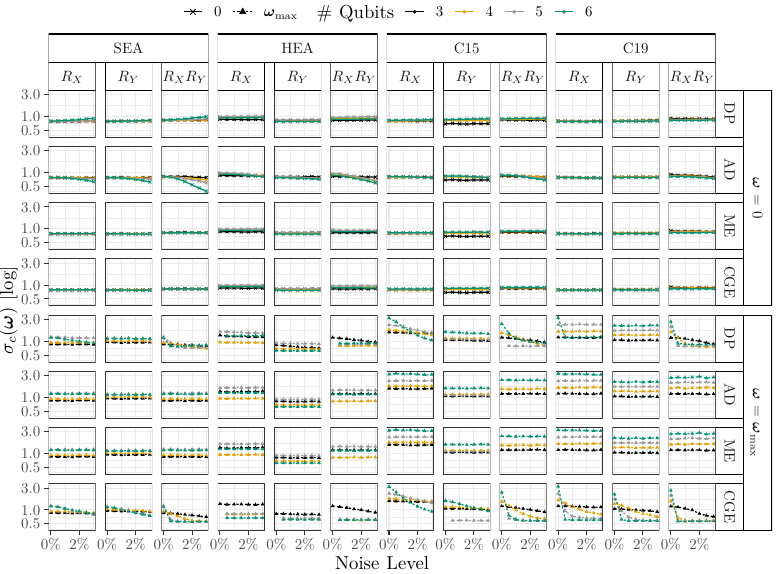}
  \caption{Relative standard deviation $\sigma_c(\bomega)$ for the absolute coefficient mean values from \autoref{fig:coefficients_mean_1D_2D}.}
  \label{fig:coefficients_sd_1D_2D}
\end{figure}

Our focus is on the effects of noise on two specific coefficient classes: the zero-frequency coefficient ($\bomega = \boldsymbol{0}$) and the coefficient corresponding to the maximum frequency in all input dimensions in the respective spectrum, determined throughout each noise type ($\bomega = \bomega_\text{max}$).
For intermediate frequencies, as illustrated in~\autoref{fig:coefficients_mean_encoding}, the coefficients decrease.
Nevertheless, the effect that noise has on these intermediate coefficient, can also be observed either at $\bomega = \boldsymbol{0}$, or $\bomega = \bomega_\text{max}$.
A selection of the full result sets, available in the supplementary material~\cite{franz_reproduction_2025}, is shown in \autoref{fig:coefficients_mean_1D_2D} and \autoref{fig:coefficients_sd_1D_2D}.

Noticeable, \gls{dp}, as a representative of the class of decoherent gate errors, uniformly causes an exponential decline of $\mu_c(\bomega)$ for most coefficients and frequencies, while for higher frequencies, mainly for higher qubit counts and for $R_X$ and $R_XR_Y$ encoding $\mu_c(\bomega)$ also increases.
This may seem counter-intuitive, but can be explained by a low coefficient mean for these frequencies without noise.
Adding noise randomly distorts the expectation value, leading to a uniform increase (on average) of all frequencies in the spectral representation.
The effect is less pronounced for lower frequency components as they typically have a higher mean value.
The relative standard deviation $\sigma_c(\bomega)$ remains constant or decreases for the special cases where $\mu_c(\bomega)$ increases.

For the zero-coefficient, we observe that applying an \gls{ad} leads to an increase in $\mu_c(\bomega)$ and a decrease in $\sigma_c(\bomega)$.
This effect is more noticeable in circuits with a higher number of qubits and input features.
While this shift is more evident in the \gls{sea} and \gls{hea}, it is less noticeable for \gls{c15} and \gls{c19}, suggesting these \ase potentially exhibit greater resilience against \gls{ad}.
For the remaining frequencies (specifically $\bomega_\text{max}$), \gls{ad} has an imperceptible effect.
Similarly, \gls{spam} errors (represented by \gls{me}), and \gls{pd}, not detailed in \autoref{fig:coefficients_mean_1D_2D} and \autoref{fig:coefficients_sd_1D_2D}, exhibit no noticeable effect for all $\bomega$.
Overall, these observations align with the proofs on decoherent noise effects in Ref.~\cite{fontana_spectral_2022}, showing (uniform) contractions for (\gls{dp}) noise channels.

The relative standard deviation $\sigma_c(\bomega)$ for $D=2$ ($R_XR_Y$ encoding), shown in \autoref{fig:coefficients_sd_1D_2D}, demonstrates similar behaviour as observed in \autoref{fig:coefficients_var} and \autoref{fig:coefficients_var_rx}, with constant or decreasing values throughout the noise levels, indicating an overall decrease of variance with $\mu_c(\bomega)$.
Regarding \glspl{cge}, their influence is minimal, when $\bomega$ is small.
However, for high frequencies, there is a significant increase in $\mu_c(\bomega)$ and decrease in $\sigma_c(\bomega)$ with increasing noise levels in configurations with more than three qubits and $D=2$.
This effect, discussed in depth in \autoref{sec:coherent_exp}, is a consequence of the uniform increase of coefficients across the full spectrum, when applying \glspl{cge}.

\section{Coefficients during Training for Problem Instances}
\label{apx:training_instances}

\begin{figure}[tbp]
  \includegraphics[width=\columnwidth]{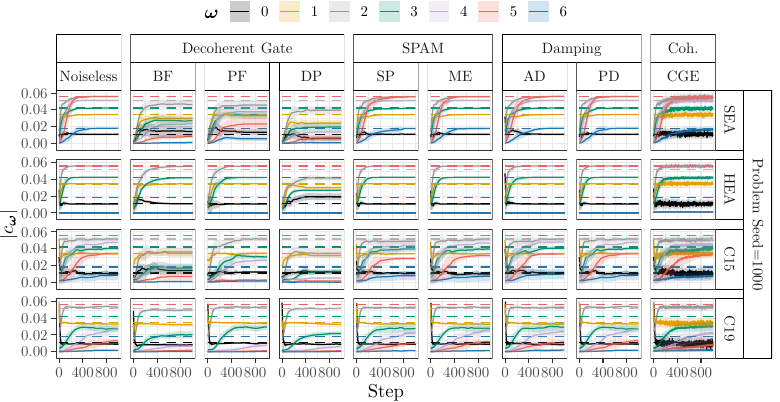}
  \caption{The learned absolute coefficients of the \glspl{qfm} averaged over ten parameter model initialisation seeds for a single randomly generated problem instances and different types of noise (3\%) during training of \num{1000} steps. Shaded areas correspond to the standard deviation across the parameter initialisation seeds from the mean. The coefficients of the objective Fourier series are marked with dashed lines.}
  \label{fig:training_seed1000}
\end{figure}

\autoref{fig:training_seed1000} shows the exact absolute coefficient values over the course of the training described in \autoref{sec:training_exp} for one of the ten seeds used for the generation of the objective functions.
The analogues for the remaining problem seeds are available in our supplementary material in Ref.~\cite{franz_reproduction_2025}.

In the noiseless scenario, the \gls{sea} is capable of learning all coefficients accordingly.
It can be observed that the higher frequencies are more prone to the effect of decoherent gate errors compared to lower frequencies, whereas \gls{spam} and \gls{dp} errors do have a negligible effect.
Interestingly, \ase which do not exhibit a full spectrum, that is the \gls{hea}, are rarely affected by errors.
The \glspl{cge} generally adds high frequent noise on each coefficient magnitude which is feasible as it is modelled by a normal distribution, adding to any gate operation, therefore resulting in white noise in the expectation value.

In the noiseless scenario, it can be observed that \eg in case of the \gls{sea}, lower frequencies are fitted faster than high-frequent coefficients which is in line with an observation made in Ref.~\cite{duffy_spectral_2025}.

\section{Entanglement during Training}
\label{apx:entanglement_training}

\begin{figure}[tbp]
  \includegraphics[width=\columnwidth]{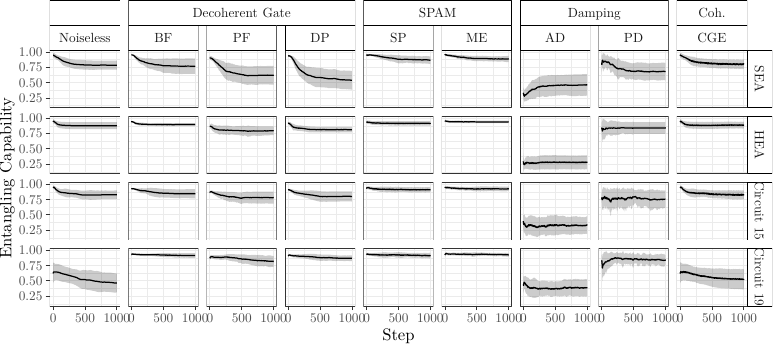}
  \caption{Entangling capability, assessed with the \gls{mw} measure for pure states (Noiseless and \gls{cge}) and \gls{ef} for mixed states (decoherent-, \gls{spam}- and damping errors) during training.
    Lines represent the mean over ten parameter initialisation seeds and ten problem generation seeds. Shaded areas represent the standard deviation.
  }
  \label{fig:entanglement_training}
\end{figure}

In \autoref{fig:entanglement_training} we show the entangling capability over the course of a training.
The orders of magnitude for the entangling capability under the influence of the aforementioned noise types are similar to those measured in \autoref{sec:entanglement}.
Over the course of the training we observe a decrease in the entangling capability for most configurations (with the exception of damping errors).
While the decrease is more pronounced in some experiments (\eg \gls{sea} with \gls{dp}), the overall trend is consistent across all tested problem and model initialisation seeds for each individually configuration.

As indicated by, for instance, the \gls{sea} in the noiseless case, which achieved an \gls{mse} near zero (\cf \autoref{fig:training_mse}) and also converges against a concrete entangling capability, there may exists an optimal degree of entanglement for a learning task in question.
However, given that the \gls{sea}, \gls{hea} and \gls{c15} obtain a similar entangling capability value, yet result in rather different \glspl{mse}, entangling capability is not a sufficient criterion on which to determine the performance of a \gls{qfm}.

\end{document}

%% file: commands.tex
\newcommand{\eg}{\emph{e.g.}\xspace}
\newcommand{\ie}{\emph{i.e.}\xspace}
\newcommand{\cf}{\emph{cf.}\xspace}

\newcommand{\giturl}{https://github.com/cirKITers/effect-of-noise-in-qfms}
\newcommand{\zenodourl}{https://doi.org/10.5281/zenodo.15211318}

\newcommand{\ketn}[1]{\vert#1\rangle^{\otimes{n}}}
\newcommand{\bran}[1]{\langle#1\vert^{\otimes{n}}}

\newcommand{\btheta}{\ensuremath \boldsymbol{\theta}}
\newcommand{\bomega}{\ensuremath \boldsymbol{\omega}}
\newcommand{\bepsilon}{\ensuremath \boldsymbol{\epsilon}}
\newcommand{\bOmega}{\ensuremath \boldsymbol{\Omega}}
\newcommand{\bx}{\ensuremath \boldsymbol{x}}
\newcommand{\be}{\ensuremath \boldsymbol{e}}
\newcommand{\I}{\ensuremath \imath} 
\newcommand{\Real}{\ensuremath \operatorname{Re}}
\newcommand{\Imag}{\ensuremath \operatorname{Im}}
\newcommand{\Cov}{\ensuremath \operatorname{Cov}}
\newcommand{\diffw}{\ensuremath \Delta c_{\bomega}}

\newcommand{\Ase}{Ansätze\xspace}
\newcommand{\ase}{ansätze\xspace}

\newcommand{\as}{ansatz\xspace}

\usepackage{pifont}
\makeatletter
\newcommand{\linebreakand}{%
  \end{@IEEEauthorhalign}
  \hfill\mbox{}\par
  \mbox{}\hfill\begin{@IEEEauthorhalign}
}
\makeatother

\definecolor{lfd1}{HTML}{FFFFFF} 
\definecolor{lfd2}{HTML}{E69F00}
\definecolor{lfd3}{HTML}{999999}
\definecolor{lfd4}{HTML}{009371}

\definecolor{lfd5}{HTML}{BEAED4}
\definecolor{lfd6}{HTML}{ED665A}
\definecolor{lfd7}{HTML}{1F78B4}

\makeatletter
\AtBeginDocument
{
  \def\ltx@label#1{\cref@label{#1}}
  \def\label@in@display@noarg#1{\cref@old@label@in@display{#1}}
  \def\label@in@mmeasure@noarg#1{%
    \begingroup%
    \measuring@false%
    \cref@old@label@in@display{#1}
    \endgroup}%
} %
\makeatother

%% file: acronyms.tex
\setabbreviationstyle[acronym]{long-short}
\glssetcategoryattribute{acronym}{nohyperfirst}{true}
\renewcommand*{\glstextformat}[1]{\textcolor{black}{#1}}

\renewcommand*{\glsdonohyperlink}[2]{%
{\glsxtrprotectlinks \glsdohypertarget{#1}{#2}}}

\newacronym{nisq}{NISQ}{Noisy Intermediate Scale Quantum}
\newacronym{qaoa}{QAOA}{Quantum Approximate Optimisation Algorithm}
\newacronym{ftqc}{FTQC}{Fault Tolerant Quantum Computing}
\newacronym{qml}{QML}{Quantum Machine Learning}
\newacronym{qfm}{QFM}{Quantum Fourier Model}
\newacronym{qc}{QC}{Quantum Computing}
\newacronym{vqc}{VQC}{Variational Quantum Circuit}
\newacronym{vqe}{VQE}{Variational Quantum Eigensolver}
\newacronym{ml}{ML}{Machine Learning}
\newacronym{spam}{SPAM}{State Preparation and Measurement}
\newacronym{sp}{SP}{State-Preparation}
\newacronym{me}{ME}{Measurement} 
\newacronym{pd}{PD}{Phase Damping}
\newacronym{bf}{BF}{Bit Flip}
\newacronym{pf}{PF}{Phase Flip}
\newacronym{ad}{AD}{Amplitude Damping}
\newacronym{dp}{DP}{Depolarisation}
\newacronym{cge}{CGE}{Coherent Gate Error}
\newacronym{kl}{KL}{Kullback-Leibler}
\newacronym{fft}{FFT}{Fast Fourier Transform}
\newacronym{rff}{RFF}{Random Fourier Feature}
\newacronym{mse}{MSE}{Mean Squared Error}
\newacronym{mw}{MW}{Meyer-Wallach}
\newacronym{ef}{EF}{Entanglement of Formation}
\newacronym{hea}{HEA}{Hardware-Efficient Ansatz}
\newacronym{sea}{SEA}{Strongly-Entangling Ansatz}
\newacronym{bp}{BP}{Barren-Plateau}
\newacronym{qpu}{QPU}{Quantum Processing Unit}
\newacronym{c15}{C15}{Circuit 15}
\newacronym{c19}{C19}{Circuit 19}